\documentclass[journal, 10pt]{IEEEtran}

 \usepackage{amsmath}
 \usepackage{amsfonts}
 \usepackage{amssymb}
\usepackage{algorithmic}
\usepackage{algorithm}
\usepackage{color}
\usepackage{graphicx}
\usepackage{longtable}
\usepackage{epstopdf}
\usepackage{url}

\makeatletter
\let\MYcaption\@makecaption
\makeatother
\usepackage[font=footnotesize]{subcaption}
\makeatletter
\let\@makecaption\MYcaption
\makeatother

\newcommand{\seclabel}[1]{\label{sec:#1}}

\newtheorem{theorem}{Theorem}[section]

\newtheorem{example}[theorem]{Example}

\let\olditemize=\itemize
\def\itemize{
\olditemize
  \setlength{\itemsep}{1pt}
  \setlength{\parskip}{0pt}
  \setlength{\parsep}{0pt}
}
\let\oldenumerate=\enumerate
\def\enumerate{
\oldenumerate
  \setlength{\itemsep}{1pt}
  \setlength{\parskip}{0pt}
  \setlength{\parsep}{0pt}
}

\renewcommand{\vec}[1]{\boldsymbol{#1}}
\DeclareMathOperator*{\var}{\text{Var}}
\DeclareMathOperator*{\cov}{\text{Cov}}
\DeclareMathOperator*{\std}{\text{Std}}

\newcommand{\eqnref}[1]{(\ref{eqn:#1})}
\newcommand{\secref}[1]{Sec. \ref{sec:#1}}
\newcommand{\appref}[1]{Appendix \ref{sec:#1}}
\newcommand{\figref}[1]{Fig. \ref{fig:#1}}
\newcommand{\tabref}[1]{Table \ref{tab:#1}}

\newcommand{\ignore}[1]{}

\newcommand{\revA}[1]{{\color{black}{#1}}}

\newcommand{\mb}[1]{\mathbf{#1}}
\newcommand{\mc}[1]{\mathcal{#1}}

\newcommand{\ie}{\textit{i.e.}}
\newcommand{\eg}{\textit{e.g.}}
\newcommand{\etc}{\textit{etc.}}
\newcommand{\aka}{\textit{a.k.a.}}

\def\ci{\perp\!\!\!\perp}

\begin{document}


\title{Multiple-Population Moment Estimation: Exploiting Inter-Population Correlation for Efficient Moment Estimation in Analog/Mixed-Signal Validation}

%
%

\author{Chenjie~Gu,~\IEEEmembership{Member,~IEEE,}
	Manzil~Zaheer,~\IEEEmembership{Student Member,~IEEE}
        and~Xin~Li,~\IEEEmembership{Senior Member,~IEEE}

}

\markboth{Manuscript}{TCAD}

\date{}

\maketitle
\begin{abstract}
Moment estimation is an important problem during circuit validation, in both pre-Silicon and post-Silicon stages. From the estimated moments, the probability of failure and parametric yield can be estimated at each circuit configuration and corner, and these metrics are used for design optimization and making product qualification decisions. 
The problem is especially difficult if only a very small sample size is allowed for measurement or simulation, as is the case for complex analog/mixed-signal circuits.
In this paper, we propose an efficient moment estimation method, called Multiple-Population Moment Estimation (MPME), that significantly improves estimation accuracy under small sample size. 
The key idea is to leverage the data collected under different corners/configurations to improve the accuracy of moment estimation at each individual corner/configuration. 
Mathematically, we employ the hierarchical Bayesian framework to exploit the underlying correlation in the data.
We apply the proposed method to several datasets including post-silicon measurements of a commercial high-speed I/O link, and demonstrate an average error reduction of up to 2$\times$, which can be equivalently translated to significant reduction of validation time and cost.
\end{abstract}

\IEEEpeerreviewmaketitle

\section{Introduction}

During circuit validation, it is crucial to make statistically valid predictions of the circuit performances of interest. The statistical nature of the problem comes from the fact that the latest process technology witnesses increasingly larger variability, and that systems are becoming so complex that effects from environment and surrounding circuits cannot be neglected, and as a result, they exhibit randomness in the circuit performance.
Such statistical predictions are important since they are used to guide design optimization and to make key decisions such as whether the product is ready for high-volume manufacturing/shipping.
A key problem in this process is the problem of estimating the probability distribution of circuit performances, \aka, density estimation.
From this distribution, metrics such as the probability of failure (PoF) or yield can be derived for further analysis and optimization.

Traditional approaches for density estimation \cite{bishop2006pattern, friedman2001elements} include parametric and non-parametric methods. 
\ignore{Parametric methods often formulate an optimization problem (\eg, maximum likelihood or maximum a posteriori) to solve for the parameters for a class of distributions. Non-parametric methods avoid the parametric assumption of the underlying distribution, and can be thought of as estimating the distribution by smoothing data. }While the existing techniques have obtained much success in various applications, they all require ``enough'' number of samples for the result to be accurate. That is, if the sample size is small, the result can be biased by the data, and may not be trusted.

This is the ``small-sample-size'' problem in circuit validation. It is further exacerbated for analog/mixed-signal applications, because both simulation and measurement of many analog/mixed-signal circuit performances are time and cost consuming \cite{casper2009, hakim2010,  guiccad2012}.
For example, post-layout simulation can be slow, especially for circuits such as SRAM/PLL where extremely small time steps are required for high accuracy.
As another example, during post-Silicon validation, due to tight product release schedules, only a limited amount of measurement may be performed within the post-Silicon time-frame.
In addition, the measurement of performance metrics, such as Bit-Error-Ratio (BER) and Time/Voltage Margins of high-speed I/O links, takes a long time, and requires expensive equipment (such as BER testers) \cite{intel1, intel2, intel3}.
Taking into consideration all the practical issues, a very small number of samples are affordable within reasonable timeframe.
\ignore{That poses serious problem for circuit validation.}

\ignore{
The small sample size makes traditional density estimation methods not applicable because they are built upon the assumption that ``enough'' data is available for valid statistical estimation. 
When the assumption is broken, we obtain low confidence in the estimated quantities. In another word, this means that we may either under-validate or over-validate the circuit.
Similar to over-design and under-design, over-validation and under-validation are as harmful, if not more, in terms of cost and time-to-market.}
Unfortunately, there is few existing satisfying solution to get around this problem. To the best of our knowledge, 
the usual practice is to increase the sample size as much as possible to reach a certain confidence level, or to set an empirical guard-band on top of the estimation.
There is a recent work \cite{bmfdac2010} that considers a similar problem, but for performance modeling.
Another recently published technique \cite{bmficcad2012} solves a similar problem for post-layout performance distribution estimation, but with mildly small number of samples (50 or more).

Another problem that is sometimes ignored in circuit validation is that circuit performance distributions need to be estimated at various corners and configurations for various similar products at different steppings.
For example, during I/O interface validation such as PCIE\cite{pcisig} and DDR\cite{jedec}, in addition to the traditional process, voltage and temperature (PVT) corners, we must also validate against different board/add-in card/Dual In-line Memory Module(DIMM) configurations, input patterns, different equalization settings, \etc.
In another word, the interface should meet the PoF specification for any customer configuration of board and add-in cards. Therefore, it is inappropriate to mix the measurements under different configurations, because even with a low PoF across all configurations, we may obtain a very high PoF at a particular configuration.
In this case, combining data from all configurations does not help us to increase the sample size. In fact,  estimating the overall distribution can lead to misleading validation results.

In this paper, we present Multiple-Population Moment Estimation (MPME) which encapsulates a class of methods to efficiently estimate the moments of performance distributions at multiple corners and configurations.
We try to solve the small sample size problem (\ie, sample size ranging from 1 to 10) by exploiting the underlying correlation of data collected at multiple corners and configurations.
In particular, we emphasize that data collected at different design stages, different configurations and different corners are not independent, but are correlated.
Taking advantage of this non-intuitive fact leads to a theoretically guaranteed better estimator.
While we focus on the moment estimation problem in this paper, it is possible to extend the idea to more general parametric and non-parametric density estimation problems.

Mathematically, MPME builds a generative graphical model to model the data obtained from simulation and measurement.
Equivalently, the statistical graphical model defines a (parameterized) joint prior distribution of the moments at multiple populations.
With the graphical model, MPME estimates the moments in two steps.
First, the Maximum Likelihood Estimation (MLE) method is used to learn the prior distribution of moments.
Second, the prior distribution learned in the first step is used to obtain the Maximum A Posteriori (MAP) estimation of moments at individual populations.
Experimental results show that in comparison to traditional sample moment estimators, 
MPME reduces the average error by up to 2x in the best case for examples obtained from measurement of commercial designs.

The rest of paper is organized as follows. 
\secref{background} formulates the problem, and explains why existing techniques can be problematic when a small number of samples are present.
\secref{method} describes rational and theory behind the MPME approach, and \secref{remark} discusses advantages, potential limitations and practical applications of the method.
\secref{examples} presents experimental results on several datasets to demonstrate that MPME is consistently superior than traditional techniques in terms of estimation accuracy.

\section{Background and Problem Formulation}
\label{sec:background}

In this paper, we consider the problem of estimating a circuit performance metric, denoted by $x$, which depends on many variables such as process parameters, voltage, temperature, board, add-in card, \etc. The performance metric $x$ can also depend (indirectly) on time, because a subset of the parameters, such as process parameters, change over time.

As a concrete example application, we consider the problem of post-Silicon validation of high speed I/O interfaces. 
In this application, a configuration is defined by fixing the values of a subset of the parameters. 
By considering variability of all the other parameters, $x$ exhibits a distribution at  each configuration.
For example, a configuration of an I/O link can be defined by the combination of a specific board and a specific add-in card.
The variability of time/voltage margin (of the eye diagram) is caused by  parameter variations such as PVT variations. 
Measurement of margins is repeated at each configuration for each Silicon stepping, and the goal of validation is to ensure that PoF meets the specification at each stepping and at each configuration.

\subsection{Problem Formulation}
\label{sec:problemformulation}

To formalize the above description, we define a \textit{population} to be a specific (corner, configuration, stepping) combination\footnote{In this definition, (VT) \textit{corner} refers to the assignment of supply voltage or temperature; \textit{configuration} refers to the I/O link configurations such as data rate, board impedance, add-in card; \textit{stepping} refers to a Silicon tape-out. Obviously, this definition is closely related to the post-Silicon I/O validation problem. Readers can define the population that suits the application at hand.},
and denote $P$ by the number of populations.
For each population, we define a random variable $x_i$, $(i=1, \cdots, P)$ to model the variability of the performance metric at the corresponding population,
and $x_i$ satisfies a Gaussian distribution $x_i \sim \mc{N}(\mu_i, \sigma_i^2)$ where $\mu_i$ is the mean and $\sigma_i^2$ is the variance.
For notational convenience, we define $\vec{\mu} = [\mu_1, \cdots, \mu_P]^T$ and $\vec{\sigma^2} = [\sigma_1^2 \cdots, \sigma_P^2]^T$. 

In this formulation, the Gaussian distribution assumption is a simplification of the problem which is often used in practice. 
We discuss the potential extensions to non-Gaussian distributions and higher-order moments in \secref{nongaussian}.

For each population, we obtain a set of independent observations $\mc{X}_i = \{ x_{i,1}, \cdots, x_{i,N_i}\}$, where $N_i$ is the sample size of the $i$-th population. Each element in $\mc{X}_i$ corresponds to one independent measurement at the $i$-th population.
The problem we aim to address is to estimate the moments $(\mu_i, \sigma_i^2), i=1, \cdots, P$, given the observations $\{\mc{X}_1, \cdots, \mc{X}_P\}$.

For example, in \secref{ioexample}, $\mc{X}_i, i=1, \cdots, 8$ represent 8 sets of observations at 8 different link configurations, and $x_{i,j}, j=1, 2, \cdots$ represent the time margin measurement of the I/O link. We would like to estimate the time margin distributions at 8 different configurations by estimating the first two moments.

The difficulty of this problem is that the sample sizes $N_i$'s can be extremely small. On the one hand, each individual sample can be very expensive to obtain due to long simulation/measurement time. On the other hand, since the validation must be performed at each configuration and corner, we have to obtain $\sum_{i=1}^P N_i$ samples in total. With a large $P$, it might be impossible to obtain that many samples within a reasonable amount of time. This effectively results in even smaller $N_i$'s.
With a very small sample size, the estimated moments could have a large error.

\subsection{Low Confidence under Small Sample Size}
\seclabel{sss}

For a specific population, the most widely used estimator for mean and variance is the sample mean $\bar{x}_i$ and sample variance $S_i$, respectively,
\begin{equation}
\begin{aligned}
\bar{x}_i = \frac{1}{N_i} \sum_{j=1}^{N_i} x_{i,j}, \quad S_i =& \frac{1}{N_i - 1} \sum_{j=1}^{N_i} (x_{i,j} - \bar{x}_i)^2.\\
\end{aligned}
\end{equation}

Since $\bar{x}_i \sim \mc{N}(\mu_i, \frac{\sigma_i^2}{N_i})$ and $S_i \sim \frac{\sigma_i^2}{N_i-1} \chi^2_{N_i-1}$,  we obtain
\begin{equation}
\std (\bar{x}_i) = \frac{1}{\sqrt{N_i}} \sigma_i, \quad \std (S_i) = \frac{\sqrt{2}}{\sqrt{N_i-1}} \sigma_i^2.
\label{eqn:stdsample}
\end{equation}

If the standard deviation of an unbiased estimator is used as a measure of accuracy and confidence level, 
\eqnref{stdsample} shows that the accuracy of both sample mean and sample variance estimators depend on $N_i$.
As $N_i$ approaches infinity, the error converges to 0. However, when $N_i$ is small, both estimators suffer from significant error.

\subsection{Handling Multiple Populations}
\label{sec:multiplepop}

One common way to handle multiple populations is to build a performance model. For example, consider the $P, V, T$ variations, one might fit a response surface model (RSM) \revA{\cite{rsmbook}}
\begin{equation}
x = h(P, V, T).
\end{equation}
Define the $i$-th population by a specific $(V, T)$ combination, denoted by $(v_i, t_i)$, we have
\begin{equation}
x_i = h(P, v_i, t_i),
\end{equation}
from which the distribution of $x_i$ can be derived given the distribution of $P$.

This is a viable solution, but its success is dependent on two critical assumptions. First, the configuration variables are continuous (not categorical).
Second, $x$ has a strong dependence on the configuration variables,
and the underlying performance model template (such as RSM) is correct.
These assumptions can often be broken in practice. 
\ignore{For example, the configuration variables, such as board and add-in card, are categorical; in addition, RSM modeling may not work well for some configuration variables due to strong nonlinearity and/or noisy measurements.}
Furthermore, a potential drawback for RSM technique is that the number of measurements must be at least as many as the number of underlying random variables. If there are too many parameters (\eg, for characterizing process variability), we need many measurements which might not be affordable.
Other techniques must to be sought to handle multiple populations.

\section{Multiple Population Moment Estimation}
\label{sec:method}

\subsection{Overview}
As is evident in \secref{sss}, if each population is treated independently, there is little room for improvement.
In contrast, MPME views data at different populations as correlated, and it tries to exploit such correlation to improve the accuracy of the estimator.

To model the correlation, MPME imposes a generative graphical model\footnote{Graphical models\cite{bishop2006pattern}, \aka, probabilistic graphical models, provide a way to describe the probabilistic structure in a set of random variables. We provide a short introduction in \appref{graphicalmodels} that is relevant for this paper.} which describes how the data are generated at multiple populations.
Equivalently, it specifies a joint prior distribution on the moments $\mu_i$'s and $\sigma_i^2$'s.
For example, the generative graphical model shown in \figref{generativemodelgaussian} specifies a model where $(\mu_i, \sigma_i^2)$ follow a distribution $p(\mu_i, \sigma_i^2|\theta)$ parameterized by $\theta$, and the $i$-th population $x_i$ follows a Gaussian distribution with mean $\mu_i$ and variance $\sigma_i^2$.

With the graphical model, MPME follows a two-step approach to estimate the moments.
\begin{itemize}
\item First, a prior distribution of $p(\mu_i, \sigma_i^2 | \theta)$  is learned from data at all populations, using Maximum Likelihood Estimation.
\item Second,  Maximum A Posteriori (MAP) estimation is applied to each population using the prior distribution learned from the first step.
\end{itemize}

\subsection{Correlation Helps Improving Estimation Accuracy}

Before we introduce MPME, it is instructive to look at two specific examples for which we can perform error analysis. The closed-form expressions intuitively explain why correlation can help improving estimation accuracy.
It can also be shown that the estimators described in the two examples can be thought of as extreme cases of MPME. In these two examples, for simplicity, we consider the case where all populations have same number of independent samples, i.e. $N_1 = \cdots = N_P = N$.

\begin{example}[unequal mean, equal variance]
$\quad$
Assume that $\mu_i$'s are different, and $\sigma_1^2 = \cdots = \sigma_P^2 = \sigma^2$, and consider the problem of estimating $\sigma^2$. 

Since $S_i \sim \frac{\sigma^2}{N-1} \chi^2_{N-1}$, we obtain an unbiased estimator for $\sigma^2$,
\begin{equation}
\frac{1}{P}[S_1 + \cdots + S_P] \sim \frac{1}{P} \frac{\sigma^2}{N-1} \chi_{NP-P}^2,
\end{equation}
from which  $\std (\frac{1}{P}[S_1 + \cdots + S_P]) = \sigma^2 \sqrt{\frac{2}{P(N-1)}}$. Hence, the estimation error decreases as $P$ increases, and is smaller than $\std(S_i)$.
\label{ex:ex1}
\end{example}

\begin{example}[equal mean, unequal variance]
$\quad$
Assume that $\mu_1 = \cdots = \mu_P = \mu$, and $\sigma_i^2$'s are different, and consider the problem of estimating $\mu$.

Since $\bar{x}_i \sim \mc{N}(\mu, \frac{\sigma_i^2}{N})$, we obtain an unbiased estimator for $\mu$,
\begin{equation}
\frac{1}{P} [\bar{x}_1 + \cdots + \bar{x}_P] \sim \mc{N} (\mu, \frac{1}{P^2} [\frac{\sigma_1^2}{N} + \cdots + \frac{\sigma_P^2}{N}]).
\label{eqn:mua}
\end{equation}
As $P$ increases, the variance of $\frac{1}{P} [\bar{x}_1 + \cdots + \bar{x}_P]$ decreases. This shows that when there are many populations, \eqnref{mua} gives a very accurate estimate of $\mu$.
\ignore{\footnote{Note, however, \eqnref{mua} is not the ``best'' estimator. Intuitively, consider an estimator of $\mu$ which is a linear combination of $\bar{x}_i$'s. Then, more weight should be given to $\bar{x}_i$ if $\sigma_i$ is smaller. However, we omit the derivation since the actual expression is a bit involved.}}
\label{ex:ex2}
\end{example}

The above two examples show that with the extra (deterministic) information of ``equal variance'' or ``equal mean'', we can reduce the estimation error roughly as $1/\sqrt{P}$.
That is, the estimation error decreases as the number of population $P$ increases. The reason for the error reduction is that the extra correlation information enables us to \textit{fuse} the data from all populations, and it effectively increases the sample size.
 
In practice, however, it is too strong a statement to claim ``equal variance'' or ``equal mean''. Rather, MPME imposes a \textit{soft} correlation structure on the mean/variance. In particular, MPME imposes a joint prior distribution $p(\vec{\mu}, \vec{\sigma^2})$ to model the correlation.

\subsection{Modeling Correlation among Multiple Populations}

\begin{figure}[h!]
  \centering
        \begin{subfigure}[b]{0.49\linewidth}
                \centering
    \includegraphics[width=0.6\linewidth]{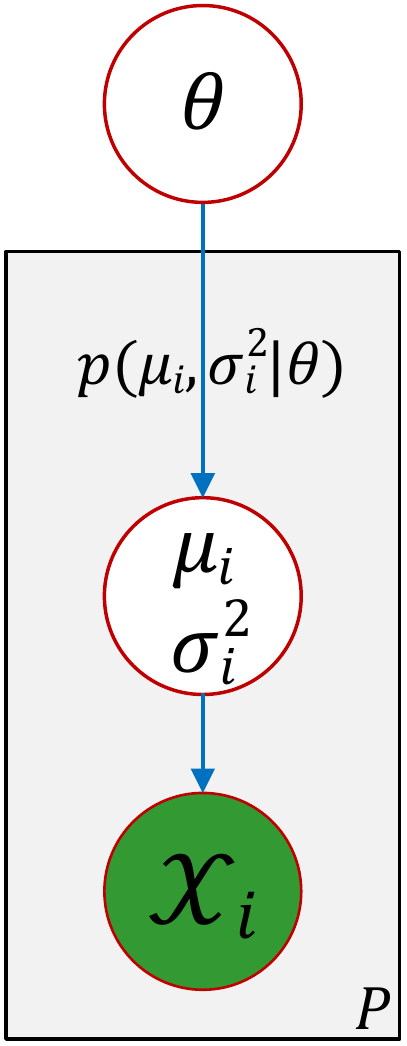}
                \caption{Probabilistic.}
                \label{fig:generativemodelgaussian}
        \end{subfigure}%
        \begin{subfigure}[b]{0.49\linewidth}
                \centering
    \includegraphics[width=0.6\linewidth]{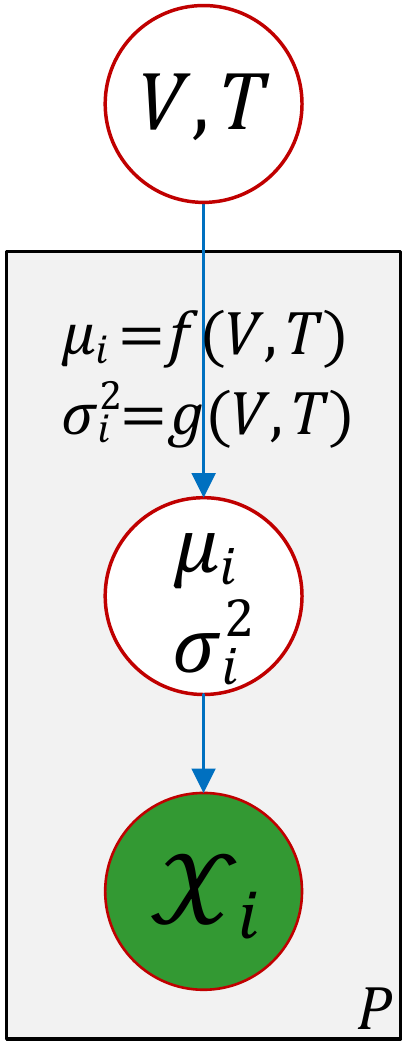}
                \caption{Deterministic.}
                \label{fig:deterministicmodel}
        \end{subfigure}
  \caption{Generative graphical models for multiple population Gaussian samples.}
\end{figure}

By imposing a joint prior distribution $p(\vec{\mu}, \vec{\sigma^2})$ on $\vec{\mu}$ and $\vec{\sigma^2}$, MPME assumes an underlying generative graphical model which describes how the data $\mc{X}_1, \cdots, \mc{X}_P$ are generated. Assuming further that the prior distribution is parameterized by $\theta$,\footnote{Here, $\theta$ is known as \textit{hyperparameters} in statistical literatures\cite{bishop2006pattern}.} the graphical model is shown in \figref{generativemodelgaussian}. 
The graphical model describes that $\mu_i$'s and $\sigma_i^2$'s are independent samples from the distribution $p(\mu, \sigma^2| \theta)$, and $\mc{X}_i$'s are conditionally independent samples from the corresponding Gaussian distributions $\mc{N}(\mu_i, \sigma_i^2)$ given $\mu_i$'s and $\sigma_i^2$'s.

Compared to the traditional approach where $\mc{X}_1, \cdots, \mc{X}_P$ are independent from each other, the graphical model in \figref{generativemodelgaussian} asserts that $\mc{X}_1, \cdots, \mc{X}_P$ are conditionally independent given $\vec{\mu}$ and $\vec{\sigma^2}$ and that $\vec{\mu}$ and $\vec{\sigma^2}$ are conditionally independent given $\theta$. Therefore, with $\theta$ unobserved, the populations $\mc{X}_1, \cdots, \mc{X}_P$ are correlated.\footnote{
We elaborate in \appref{priorcorr} the correlation induced by applying a (unobserved) prior distribution, and its relationship to traditional concept of the correlation coefficient.}
This is a key difference between the traditional approach and MPME -- it allows MPME to fuse the data from all populations, thus improving estimation accuracy.

It is important to note that in practice, the moments $\vec{\mu}$ and $\vec{\sigma^2}$ are deterministic fixed quantities given the circuit and the configuration, and are \textit{not} random variables.
For example, considering only V, T dependencies, the $\mu_i$'s and $\sigma_i^2$'s are deterministic functions of $V, T$, as shown in \figref{deterministicmodel}.
The probabilistic generative model in \figref{generativemodelgaussian} is simply a way to avoid estimating the potentially highly nonlinear functions $f(\cdot)$ and $g(\cdot)$. It replaces the deterministic function of $\mu_i$'s and $\sigma_i^2$'s with a joint distribution that approximates the correlation defined by $f(\cdot)$ and $g(\cdot)$.
However, this is a very mild assumption. The probabilistic modeling not only boosts estimation accuracy, but also provides significant scalability/flexibility compared to direct performance modeling of $\mu_i$'s and $\sigma_i^2$'s.

\ignore{
Another way to understand this formulation is that the probabilistic graphical model in fact models the correlation that is encoded in the deterministic model.
To see that, consider two populations $x_1$ and $x_2$, both of which depends on $V$, $T$ variations linearly, \ie
\begin{equation}
\begin{aligned}
x_1 =& a_1 \Delta V + a_2 \Delta T + a_3 e \\
x_2 =& b_1 \Delta V + b_2 \Delta T + b_3 e\\
\end{aligned}
\end{equation}
where $e$ models all variations except for $V, T$ variation, and
$\Delta V$, $\Delta T$ and $e$ all follow independent standard Gaussian distributions $\mc{N}(0,1)$.
Obviously,
\begin{equation}
\begin{aligned}
\var(x_1) =& a_1^2 + a_2^2 + a_3^2 \\
\var(x_2) =& b_1^2 + b_2^2 + b_3^2 \\
\cov(x_1, x_2) =& a_1 b_1 + a_2 b_2 + a_3 b_3\\
\rho_{x_1, x_2} =& \frac{\cov(x_1, x_2)}{\sqrt{\var(x_1) \var(x_2)}}\\
\end{aligned}
\end{equation}

It is not hard to see that when $e$ is dominant, \ie, $a_3 \gg a_1, a_2$ and $b_3 \gg b_1,  b_2$, then $\rho_{x_1, x_2} \to 1$, and therefore the two populations become correlated. MPME models such correlation implicitly implied by the deterministic model.
}

\begin{figure}[h!]
  \centering
    \includegraphics[width=0.294\linewidth]{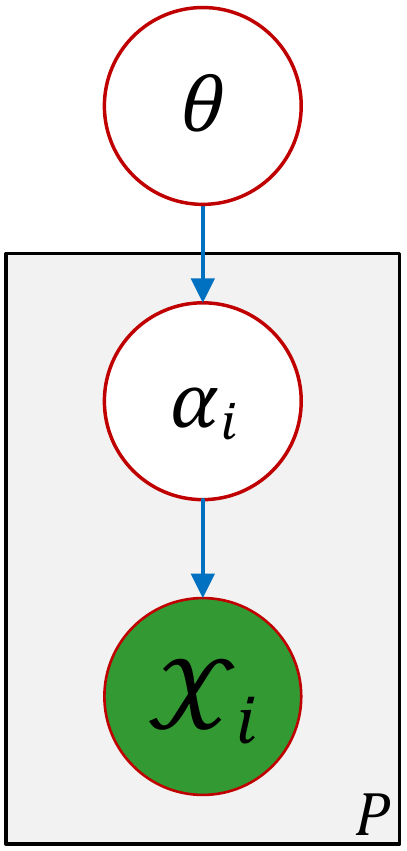}
  \caption{Generative graphical model for multiple population parametric density estimation problem.}
\label{fig:generalgenerativemodel}
\end{figure}

The above generative graphical modeling idea can be extended to more general scenarios, including parametric and non-parametric multiple population density estimation problems. For example, consider the parametric density estimation problem where $x_i$ satisfies the distribution $p(x_i | \alpha_i)$ parameterized by $\alpha_i$. By imposing a joint distribution $p(\vec{\alpha}|\theta)$ over $\alpha_i$'s, we obtain the generative graphical model in \figref{generalgenerativemodel}. The 2-step approach in MPME can be similarly applied to this model for estimating $\alpha_i$'s. However, this is out of the scope of the paper, and we will only focus only on the moment estimation problem.

\subsection{Choosing Prior Distributions}
\seclabel{priorchoice}

Intuitively, the prior distribution for $\mu_i$'s and $\sigma_i^2$'s, denoted by $p(\mu_i, \sigma_i^2)$, describes the \textit{belief} about the correlation among $\mu_i$'s and $\sigma_i^2$'s. It is useful to note that the probabilistic models encompass deterministic relationships between parameters at different populations.
For example, in Example \ref{ex:ex1}, $\sigma_1^2 = \cdots = \sigma_P^2$ corresponds to a Dirac distribution $p(\sigma_i^2) = \delta(\sigma_i^2 -\sigma^2)$, and in Example \ref{ex:ex2}, $\mu_1 = \cdots = \mu_P$ corresponds to a Dirac distribution $p(\mu_i) = \delta(\mu_i -\mu)$.

However, in real applications, it is too strong to claim a priori that $\mu_i$'s and $\sigma_i^2$'s at all populations are the same.
Instead, it is often the case that $\mu_i$'s and $\sigma_i^2$'s at different populations are similar, but not equal -- this is often observed in practical analog/mixed-signal circuits, especially in those carefully designed to account for variability. For example, many circuits have compensation loops and self-reconfigurable/self-healing features that cancel out the effects due to certain variability, which effectively pushes $\mu_i$'s towards each other.
On the other hand, the variance in the circuit performance is usually caused by a small set of parameters (such as critical process parameters, temperature, voltage), and the dependency at different configurations tends to be similar, which effectively pushes $\sigma_i^2$ towards each other. 

Based on the above observation, we consider two candidates for the prior distribution.
\ignore{\footnote{However, there is no limitation to use other prior distributions in MPME. One just need to ensure that the prior distribution roughly reflect the relationships of $\mu_i$'s and $\sigma_i^2$'s in reality, in order for the method to work well.}
}

\subsubsection{Independent Uniform Prior (UNI)}
The first candidate is the uniform prior distribution defined by
\begin{equation}
p(\mu_i, \sigma_i^2) = p(\mu_i|a, b) p(\sigma_i^2|c, d),
\label{eqn:uniformprior}
\end{equation}
where
\begin{equation}
\begin{aligned}
p(\mu_i|a, b) &=  \left\{ 
  \begin{array}{l l}
	\frac{1}{b-a} & \quad \text{if $\mu_i \in [a, b]$}\\
	0 &  \quad \text{otherwise}\\
  \end{array} \right. , \\
p(\sigma_i^2|c, d) &=  \left\{ 
  \begin{array}{l l}
	\frac{1}{d-c} & \quad \text{if $\sigma_i^2 \in [c, d]$}\\
	0 &  \quad \text{otherwise}\\
  \end{array} \right. , \\
\end{aligned}
\end{equation}
where $a, b\in \mb{R}, c, d \in \mb{R}^{+}$ are hyperparameters that satisfy $a\le b, c\le d$. 

\begin{figure}[h!]
  \centering
    \includegraphics[width=0.95\linewidth]{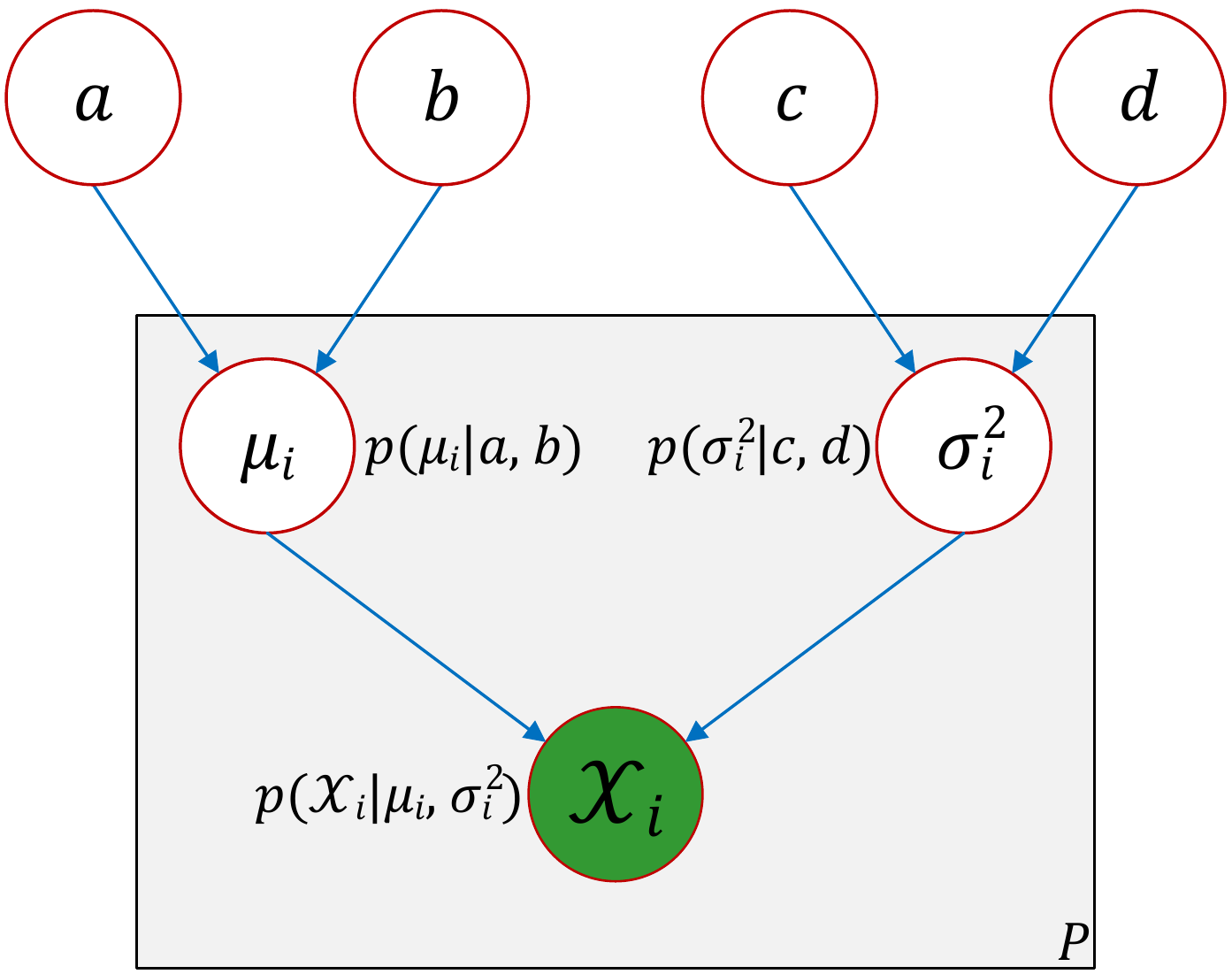}
  \caption{Generative graphical model corresponding to uniform prior (UNI).}
\label{fig:uniform_prior}
\end{figure}

As is evident from \eqnref{uniformprior}, $\mu_i$ and $\sigma_i^2$ are independent, and are parameterized by $(a, b)$ and $(c,d)$, respectively.
The corresponding generative graphical model is shown in \figref{uniform_prior}. 

The uniform prior is interesting because it has a straightforward interpretation when applied -- the process of learning a uniform prior can be thought of as obtaining a bound on the quantities to be estimated, and the process of applying the uniform prior during estimation can be thought of as restricting the estimators to be within the bound defined by $(a, b)$ and $(c, d)$.
Details of the derivation are presented in \appref{unidetails}.

\subsubsection{Normal-Inverse-Chi-Squared Prior (NIX)}
\seclabel{secnix}
The second candidate is known as the normal-inverse-chi-squared prior defined by
\begin{equation}
p(\mu_i, \sigma_i^2) = p(\mu|\sigma_i^2) p(\sigma_i^2),
\end{equation}
where
\begin{equation}
\begin{aligned}
p(\mu_i | \sigma_i^2) &= \mc{N}(\mu_i | \mu_0, \sigma_i^2 / \kappa_0), \\
p(\sigma_i^2) &= \chi^{-2}(\sigma_i^2 | \nu_0, \sigma_0^2), \\
\end{aligned}
\end{equation}
where $\mu_0 \in \mb{R}, \nu_0, \kappa_0, \sigma_0^2 \in \mb{R}^{+}$ are hyperparameters.

\begin{figure}[h!]
  \centering
    \includegraphics[width=0.95\linewidth]{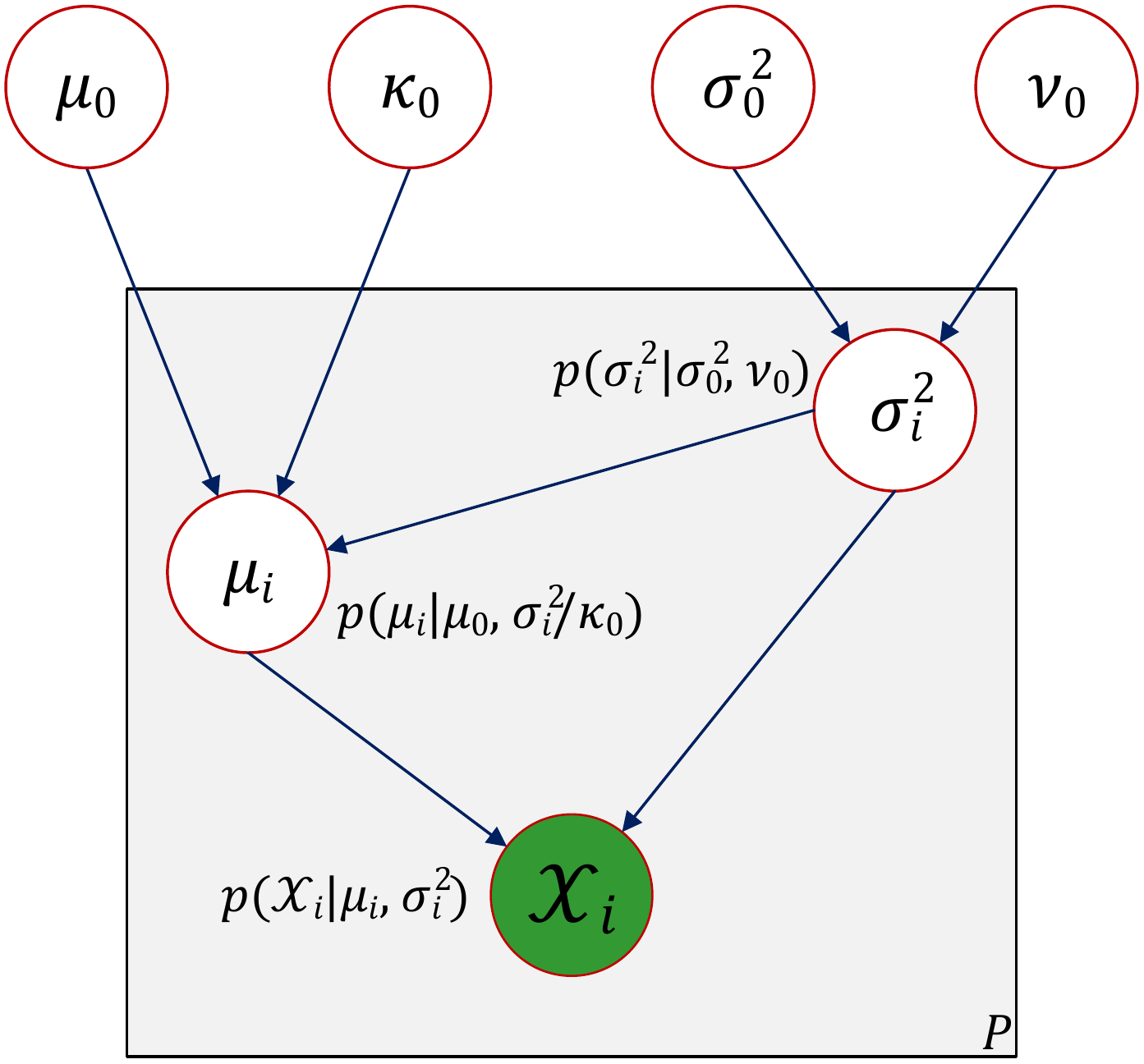}
  \caption{Generative graphical model corresponding to normal-inverse-chi-squared prior (NIX).}
\label{fig:normal_inv_chis_prior}
\end{figure}

Unlike the independent uniform prior, $\mu_i$ and $\sigma_i^2$ are not independent in the normal-inverse-chi-squared prior.
The corresponding generative graphical model is shown in \figref{normal_inv_chis_prior}.

The normal-inverse-chi-squared prior is particularly useful because it is a \textit{conjugate prior} -- \ie, the posterior distribution $p(\mu_i, \sigma_i^2 | \mc{X}_i)$ is also a normal-inverse-chi-squared distribution. It allows for closed-form expressions of the posterior, leading to closed-form expressions of the MAP solution. Therefore, the MAP estimation using this prior is extremely computationally efficient. Details of the derivation are presented in \appref{nixdetails}.

Similar to the UNI prior, the NIX prior also has a straightforward interpretation -- it is equivalent to increasing the effective number of samples by adding ``fake'' data samples that reflect the prior. As is shown in \appref{nixdetails}, the MAP mean estimation is equivalent to adding $\kappa_0$ data samples with mean $\mu_0$, and the MAP variance estimation is equivalent to adding $\nu_0$ data samples with variance $\sigma_0^2$.
Therefore, if $\kappa_0$ and $\nu_0$ are large, we effectively have more samples, and that lead to more accurate estimation.
As will be illustrated on a dataset in \secref{ex1}, MPME can significantly increase the number of effective samples.

It is also interesting to note that both prior distributions can converge to the Dirac distribution $p(\sigma_i^2) = \delta(\sigma_i^2 - \sigma^2)$ in Example \ref{ex:ex1}
and $p(\mu_i) = \delta (\mu_i - \mu)$  in in Example \ref{ex:ex2}.
For the uniform prior, the Dirac prior may be obtained as $|b-a|\to 0$ and $|d-c|\to 0$.
For the normal-inverse-chi-squared prior, the Dirac prior may be obtained as $\kappa_0 \to \infty$ and $\nu_0 \to \infty$.

\subsection{Learning the Prior Distribution}
\label{sec:learnprior}
In MPME, the first step is to learn a prior distribution from data collected at all populations. 
We employ the maximum likelihood approach to learn the prior $p(\mu_i, \sigma_i | \vec{\theta})$, where $\vec{\theta}$ are \textit{hyper-parameters} of the prior distribution.
For example, $\vec{\theta} = [a, b, c, d]$ for the UNI prior, and $\vec{\theta} = [\kappa_0, \mu_0, \nu_0, \sigma_0^2]$ for the NIX prior.

The optimization problem can be formulated as
\begin{equation}
      \begin{aligned}
        & \underset{\vec{\theta}}{\text{maximize}} & & p(\mc{X}_1, \cdots, \mc{X}_P | \vec{\theta}),\\
      \end{aligned}
      \label{eqn:mlehyperparameter}
\end{equation}
where $p(\mc{X}_1, \cdots \mc{X}_P | \vec{\theta})$ is the likelihood function. We may either use a nonlinear optimizer to solve for the optimal $\vec{\theta}$, or we may derive closed-form solutions by solving 
\begin{equation}
\frac{d}{d \vec{\theta}} p(\mc{X}_1, \cdots, \mc{X}_P | \vec{\theta}) = 0.
\end{equation}

To compute the likelihood function $p(\mc{X}_1, \cdots \mc{X}_P | \vec{\theta})$, we resort to the graphical model and integrate out $\vec{\mu}$ and $\vec{\sigma}$, \ie
\begin{equation}
\begin{aligned}
& p(\mc{X}_1, \cdots, \mc{X}_P | \vec{\theta}) \\
=& \int_{\vec{\mu}, \vec{\sigma}} p(\mc{X}_1, \cdots, \mc{X}_P | \vec{\mu}, \vec{\sigma}^2) p(\vec{\mu}, \vec{\sigma}^2|\vec{\theta}) d\vec{\mu} d\vec{\sigma}.\\ 
\end{aligned}
\label{eqn:likelihood0}
\end{equation}
The integral \eqnref{likelihood0} can be computed by numerical integration, or we may derive its closed-form expression for special prior distributions.
The derivations of $p(\mc{X}_1, \cdots, \mc{X}_P|\vec{\theta})$ for the UNI prior and NIX prior are presented in \appref{unidetails} and \appref{nixdetails}, respectively.

\ignore{shall we add a subsection discussing prior model selection? Do we have any result on that?}

\subsection{Maximum A Posteriori Estimation of $\vec{\mu}$ and $\vec{\sigma}$}
\label{sec:mapest}
Once the prior $p(\mu_i, \sigma_i^2| \vec{\theta})$ is learned, MAP estimation can be applied to obtain a point estimate of $\mu_i$'s and $\sigma_i^2$'s.
MAP formulation searches for the values of $\mu_i$'s and $\sigma_i^2$'s that maximize the posterior distribution, \ie, it solves 
\begin{equation}
      \begin{aligned}
        & \underset{\mu_i, \sigma_i^2}{\text{maximize}}  & & p(\mu_i,\sigma_i^2  |\mc{X}_i, \vec{\theta}).\\
      \end{aligned}
      \label{eqn:map}
\end{equation}

According to Bayes' rule,
\begin{equation}
p(\mu_i, \sigma_i^2 | \mc{X}_i, \vec{\theta}) \propto  p(\mc{X}_i | \mu_i,\sigma_i^2) p (\mu_i,\sigma_i^2 | \vec{\theta}),
\label{eqn:post1}
\end{equation}
where $p(\mu_i, \sigma_i^2|\vec{\theta})$ is learned as described in \secref{learnprior}, and
\begin{equation}
\begin{aligned}
&p(\mc{X}_i | \mu_i, \sigma_i^2)\\
=& \prod_{j=1}^{N_i} \frac{1}{\sqrt{2\pi\sigma_i^2}} \exp \left\{ - \frac{(x_{i,j} - \mu_i)^2}{2\sigma_i^2} \right\}\\
=&\frac{\sigma_i^{-N_i}}{(2\pi)^{N_i/2}}\exp \left\{ -\frac{N_i(\bar{x}_i-\mu_i)^2+(N_i-1)S_i}{2\sigma_i^2} \right\},\\
\end{aligned}
\label{eqn:px1}
\end{equation}
because $x_{i,j}, j=1, \cdots, N_i$ are independent samples from the Gaussian distribution $\mc{N}(\mu_i, \sigma_i^2)$.

The details of the MAP estimation for the UNI prior and NIX prior can be found in \appref{unidetails} and \appref{nixdetails}, respectively.

\subsection{MPME Algorithm}
Summarizing \secref{learnprior} and \secref{mapest}, the MPME algorithm is shown in Algorithm \ref{alg:mpme}.
\begin{algorithm}[h]
\caption{Multiple Population Moment Estimation}
\label{alg:mpme}
\textbf{Inputs:} $\mc{X}_1, \cdots, \mc{X}_P$.

\textbf{Outputs:} $(\mu_i, \sigma_i^2)$, $i=1, \cdots, P$.

\begin{algorithmic}[1]
\STATE Solve $\underset{\vec{\theta}}{\text{maximize}}\,\, p(\mc{X}_1, \cdots, \mc{X}_P | \vec{\theta})$ (\eqnref{mlehyperparameter}) for $\vec{\theta}$
\FOR{$i=1\to P$}
\STATE Solve  $\underset{\mu_i, \sigma_i^2}{\text{maximize}}\,\,  p(\mu_i,\sigma_i^2  |\mc{X}_i)$ (\eqnref{map}) for $(\mu_i, \sigma_i^2)$
\ENDFOR
\end{algorithmic}
\end{algorithm}

\section{Remarks}
\label{sec:remark}

\subsection{Practical Implementation}
\label{sec:impl}
It should be noted that the optimization problems in Algorithm \ref{alg:mpme} may not be convex, and may have multiple local optimal points. There is no guarantee that the numerical algorithm will find the global optima.
However, since initial guesses can be estimated from the same data, the optimizer has a good guess to start with, and is less affected by local optimal points.

To alleviate the computational cost associated with solving the optimization problems, we may impose an empirical prior distribution, instead of learning one from data.
For example, experienced designers may have a good idea of the range of $\sigma_i^2$ at each population (\eg, either from results of test chips or previous products) -- in this case, a uniform prior for $\sigma_i^2$'s can be asserted. However, empirical priors should be used with great caution, since it may incur unexpected bias. To be less biased, one may apply cross-validation \cite{crossvalidation1} to check the validity of the empirical prior.

\subsection{Connections to Empirical Bayes Estimators}

The ideas presented in this paper are similar to the philosophy of a class of Bayesian estimators, called \textit{Empirical Bayes estimators} (EB)\cite{eb1}. EB applies Bayes' rule to obtain either a point estimation or a posterior distribution of the parameters to be estimated. Unlike standard Bayesian methods that specify an arbitrary prior, EB learns the prior distribution from data.
In particular, if a Gaussian prior is used for the mean, EB gives the so-called \textit{James-Stein estimator}\cite{stein1} for the mean.

Particularly, a nice feature of the James-Stein estimator is that it is ``superior'' to the sample mean estimate, in the sense that the expected sum of mean square error of $\mu_i$'s at all populations is smaller than that of the sample mean estimator, \ie
\begin{equation}
E \{ \sum_{i=1}^P (\mu_i - \mu_i^{JS})^2 \} < E\{ \sum_{i=1}^P (\mu_i - \bar{x}_i)^2 \},
\label{eqn:js1}
\end{equation}
where $\mu_i$ is the actual mean, $\mu_i^{JS}$ is the James-Stein estimator and $\bar{x}_i$ is the sample mean.
One can show that if the Gaussian prior on $\mu_i$'s is used in our method, we obtain an estimator very similar to the James-Stein estimator, and \eqnref{js1} still holds.

Unlike the James-Stein estimator, our method allows for more general prior distributions. In particular, we have derived the case for the UNI prior and the NIX prior.
We will show in \secref{examples} that our method can significantly out-perform sample mean/variance estimators.

\subsection{Other Prior Distributions}
\label{sec:otherprior}
The choice of the prior distribution largely depends on its modeling capability as well as the computational tractability.
In terms of the modeling capability, both UNI and NIX prior can model the closeness of mean/variance across populations pretty well.
On the other hand, the likelihood functions using both priors have (semi-)analytical expressions, and the
MAP estimation for both priors are extremely efficient due to the simplicity of the priors.

In addition to the UNI and NIX prior mentioned in \secref{priorchoice}, one can apply other prior distributions and follow the same procedure of MPME.
Different prior distributions encode different information and therefore encourage solutions of particular structures. For example, the Laplace distribution is a prior that encourages sparsity in the solution \cite{bishop2006pattern}. More generally, one can use a mixture of Gaussian to approximate any distribution to arbitrarily accurately.
In this paper, we exploit the underlying structure that the mean/variance values cluster together, and find UNI and NIX prior are good enough for that purpose.

Using more complicated prior distributions also raises the question of computational tractability. For example, consider the problem of MAP estimation for the mean value of a Gaussian distribution. If we use a mixture of two Gaussians as the prior for the mean, then the posterior distribution for the mean is again a mixture of two Gaussians. The MAP estimation is in general no longer a convex optimization problem (as is the case for UNI and NIX priors), and therefore we lose the theoretical tractability for the MAP estimation.
In addition, it is not hard to see that the parameter learning problem will become more complicated and computationally more expensive.

\subsection{Non-Gaussian Distributions and Higher-Order Moments}
\label{sec:nongaussian}

The discussion in \secref{method} focuses on the case where the distribution at each population is Gaussian. This is an engineering assumption that is often used in practice. And with very few samples (\eg, 5), it is impossible to obtain an accurate estimation of the moments/distributions without extra knowledge about the problem.

For non-Gaussian distributions, the distributions of mean/variance are not Gaussian and $\chi^2$, and therefore the derivations need to be modified. 
The shape of the mean/variance distribution, however, may not have a closed form expression, and need to be treated on a case-by-case basis.

On the other hand, it is straightforward to extend MPME to non-Gaussian distributions if they have a limited number of parameters or sufficient statistic (\eg, the exponential family). For many distributions in the exponential family, the sufficient statistic include the first two moments of $x$ or $\ln x$. The adaption of MPME to these distributions include the choice of prior and the derivation of the posterior distribution. This is relatively straightforward because it is well established \cite{bishop2006pattern} that all members of the exponential family have conjugate priors (which lead to closed-form posterior distributions and make the MAP estimation procedure efficient).

In rare cases in circuit validation, one might also want to estimate higher-order moments (such as skewness and kurtosis). 
\revA{In theory, MPME may be applied to estimate higher-order moments, but with small sample sizes, the estimation error may still be too large for the method to be practical.}
Indeed, to apply MPME, one needs to further define $p(x|m_i)$ where $m_i$ is the $i$-th moment. The rest of the algorithm can be derived by following the steps in \secref{method}.
This means that we need a way to convert a series of moments $m_1, m_2, \cdots$ to a probability distribution. 
\revA{This is a very hard problem, and deserves a paper by itself --  \cite{apex} proposed a solution which might be used together with the MPME algorithm.}

An engineering solution to this problem is to assert $p(x^i |m_i)$ is Gaussian with mean $m_i$. In this case, MPME can be readily applied by treating $x^i$ as samples. We have used this method for the variance estimation problem and compared it to the rigorous treatment using $\chi^2$ distribution. The empirical result shows that this method is not too much worse than the rigorous method.

\subsection{Potential Limitations}
\label{sec:limitation}
Although our method may obtain a theoretically better overall estimate according to conclusions such as \eqnref{js1}, it can be the case, theoretically, that for a specific population, our method introduces a large bias.

As an extreme example, consider 100 populations, each with 1 observation, and $\mu_1 = \cdots = \mu_{99} = 0, \mu_{100} = 1$, $\sigma_1^2 = \cdots = \sigma_{100}^2 = 1$.Effectively, our method will shrink the estimated mean towards 0. Therefore, for the 100-th population, the bias can be large.

However, due to the reasons mentioned in \secref{priorchoice}, such extremely pathological cases are unlikely to happen. Even if it happens, the outliers can be easily identified in a pre-processing step, and therefore accuracy will not be compromised by outliers.

\subsection{General Guideline of Applying MPME}
There are two key questions that one may ask before applying MPME:
\begin{enumerate}
\item When is MPME (significantly) better than the sample estimators?
\item Which prior (NIX or UNI) should be used in MPME?
\end{enumerate}

While it is hard to give a definite answer and rigorous theoretical analysis, we provide several general guidelines that help answering these questions.

First, MPME is significantly better than sample estimators only if the sample size is small.
From \eqnref{stdsample}, the error of sample estimators decreases as $N$ increases. Therefore, if the sample size is large, sample estimators are good enough, and the benefit brought by MPME is negligible.

Second, MPME is significantly better than sample estimators only if the variance is large.
Similarly, from \eqnref{stdsample}, the error of sample estimators decreases as the variance $\sigma_i^2$ decreases. Therefore, if $\sigma_i$'s are small, sample estimators also give very accurate results, and MPME estimation will be very similar to that of the sample estimators.

Third, obvious outliers need to be pruned in MPME. When using MPME, it is helpful to first inspect how the sample mean/variance spread. If there are obvious population outliers, they need to be removed. As explained in \secref{limitation}, the outliers are unlikely to be correlated to other populations, and therefore including them in MPME could lead to worse results.

Fourth, empirically, the NIX prior is usually better for the overall error than the UNI prior, and the UNI prior is more consistent across populations than the NIX prior.
We can explain this empirical result by inspecting the MAP estimation equations of mean for the UNI prior and NIX prior.

Equation \eqnref{mapsigma2} shows that NIX prior pulls the mean estimate towards the prior mean $\mu_0$, which is likely to be close to the mean across all populations.
Therefore, for a specific population, if $\mu_i$  is close to the overall mean $\mu_0$, then NIX prior will give almost perfect estimation. However, the NIX prior can lead to large bias if $\mu_i$ is far from $\mu_0$.

In contrast, equation \eqnref{mapmeanuni} shows that applying the UNI prior is equivalent to applying a bound $[a,b]$ on the sample estimator.
Since $a,b$ are learned from data, they usually cover the range of the mean values of every population. This means that no matter where the $\mu_i$ is, the accuracy improvement tend to be similar because the probability that sample mean is out of the range of $[a, b]$ is low.

\ignore{
\subsection{Full Bayesian Treatment}

In our formulation, the graphical model defines a joint distribution over $(\vec{\mu}, \vec{\sigma^2}, \mc{X}_1, \cdots, \mc{X}_P, \vec{\theta})$, where $\mc{X}_1, \cdots, \mc{X}_P$ are observed.

In MPME, we first compute $\vec{\theta}^*$ by maximizing $\int_{\vec{\mu}, \vec{\sigma^2}} p(\vec{\mu}, \vec{\sigma^2}, \vec{\theta})$, \ie, the marginal distribution of $\vec{\theta}$. We then compute the most likely $\vec{\mu}$ and $\vec{\sigma^2}$ using $\vec{\theta}^*$.

A full Bayesian treatment, however, will marginalize $\vec{\theta}$ to obtain $\int_{\vec{\theta}} p(\vec{\mu}, \vec{\sigma^2}, \vec{\theta})$ (possibly using a non-informative prior of $\vec{\theta}$?), and then compute the maximum likely $\vec{\mu}$ and $\vec{\sigma^2}$.

Also, one may find $(\vec{\theta}, \vec{\mu}, \vec{\sigma^2})$ that jointly maximizes $p(\vec{\mu}, \vec{\sigma^2}, \vec{\theta})$.
(\eg, using EM algorithm to update $\vec{\theta}$ and $(\vec{\mu}, \vec{\sigma^2})$ iteratively).

How to these three options compare?
}

\section{Experimental Results}
\label{sec:examples}

In this section, we illustrate the proposed method, MPME, on a few synthetic examples as well as an industrial example of a commercial high-speed I/O link.
By the synthetic examples, we demonstrate that MPME can achieve much more accuracy compared to traditional methods such as sample mean and sample variance estimator, and we conclude empirically the scenarios under which MPME may significantly outperform traditional methods. By the industrial example, we illustrate that MPME can increase validation quality and potentially reduce test time by more than 2X.
All the numerical experiments are carried out using multiple threads on a Linux machine with Intel Xeon E5-2430 CPUs capable of running 24 threads in parallel and 64 GB of total physical memory.

\subsection{Synthetic Examples 1}
\seclabel{ex1}

In this example, the data is generated as follows:
\begin{enumerate}
\item Determine $P$ (the number of populations) and $M$ (the number of independent trials).
\item Choose $N_1 = \cdots = N_P = N$ (i.e., all populations have same number of independent samples) and determine $N$ (the number of samples at each population).
\item Choose $\mu_i$'s to be equally spaced over $[9.5, 10.5]$, \ie, $\mu_i = 9.5 + \frac{1}{P-1} (i-1)$, $i=1, \cdots, P$.
\item Choose $\sigma_i$'s to be equally spaced over $[0.95, 1.05]$, \ie, $\sigma_i = 0.95 + \frac{0.1}{P-1} (i-1)$, $i=1, \cdots, P$.
\item For $i=1, \cdots, P$, draw $x_{i,j}, j = 1, \cdots, N$ from $\mc{N}(\mu_i, \sigma_i^2)$. 
\label{stepsample}
\end{enumerate}

In our experiments, we choose $M = 500$, \ie, we generate 500 independent random trials from the same distribution. To compare MPME against sample estimators, we compute the average  error across populations, defined by
\begin{equation}
\begin{aligned}
\epsilon_{\mu} &= \frac{1}{P} \sum_{i=1}^P \sqrt{ \frac{1}{M} \sum_{j=1}^M  (\mu_i - \hat{\mu}_{i, j})^2}, \\
\epsilon_{\sigma^2} &=  \frac{1}{P} \sum_{i=1}^P \sqrt{ \frac{1}{M} \sum_{j=1}^M  (\sigma_i^2 - \hat{\sigma}^2_{i, j})^2}, \\
\end{aligned}
\end{equation}
where $\hat{\mu}_{i,j}$ and $\hat{\sigma^2}_{i,j}$ are the estimated mean/variance for the $i$-th population in the $j$-th trial.

We apply three methods (sample estimator, MPME with UNI prior and MPME with NIX prior) to this data set with varying $P$ and $N$ chosen from $P \in \{5, 10, 15, 20, 30, 40, 50, 100\}$ and $N \in \{5, 11, 15, 21, 31, 41, 51, 101\}$. 
Under all combinations of $P$ and $N$, we observe that MPME always out-performs the sample estimator in terms of accuracy.

Out of all combinations of $P$ and $N$, we discuss the results of two special cases $P=20$ and $N=5$,
to illustrate how the accuracy of MPME estimation improves with the number of samples $N$ and the number of populations $P$.

\begin{figure}[h!]
  \centering
        \begin{subfigure}[b]{0.8\linewidth}
                \centering
    \includegraphics[width=\linewidth]{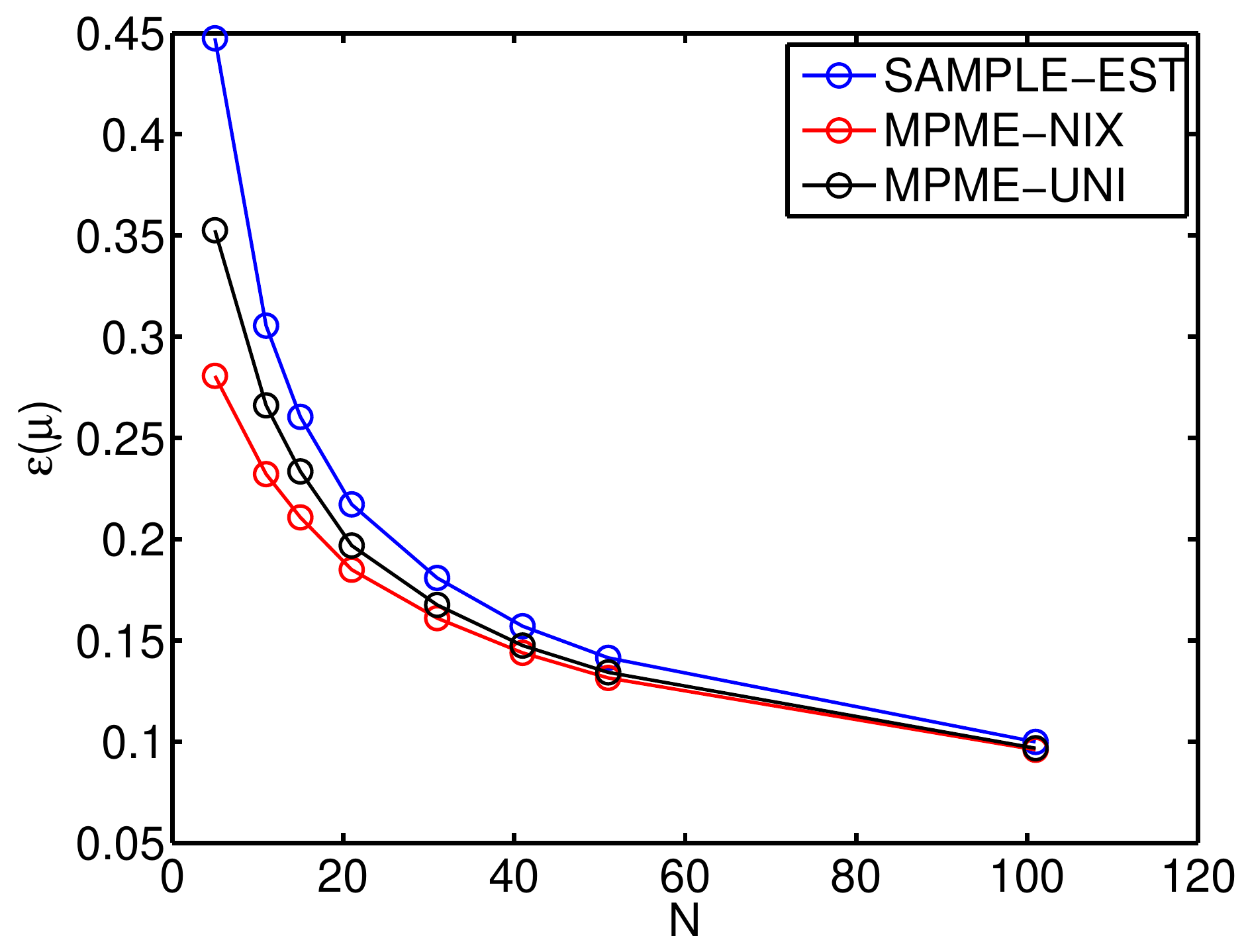}
                \caption{$\epsilon_{\mu}$ vs $N$.}
        \end{subfigure}%

        \begin{subfigure}[b]{0.8\linewidth}
                \centering
    \includegraphics[width=\linewidth]{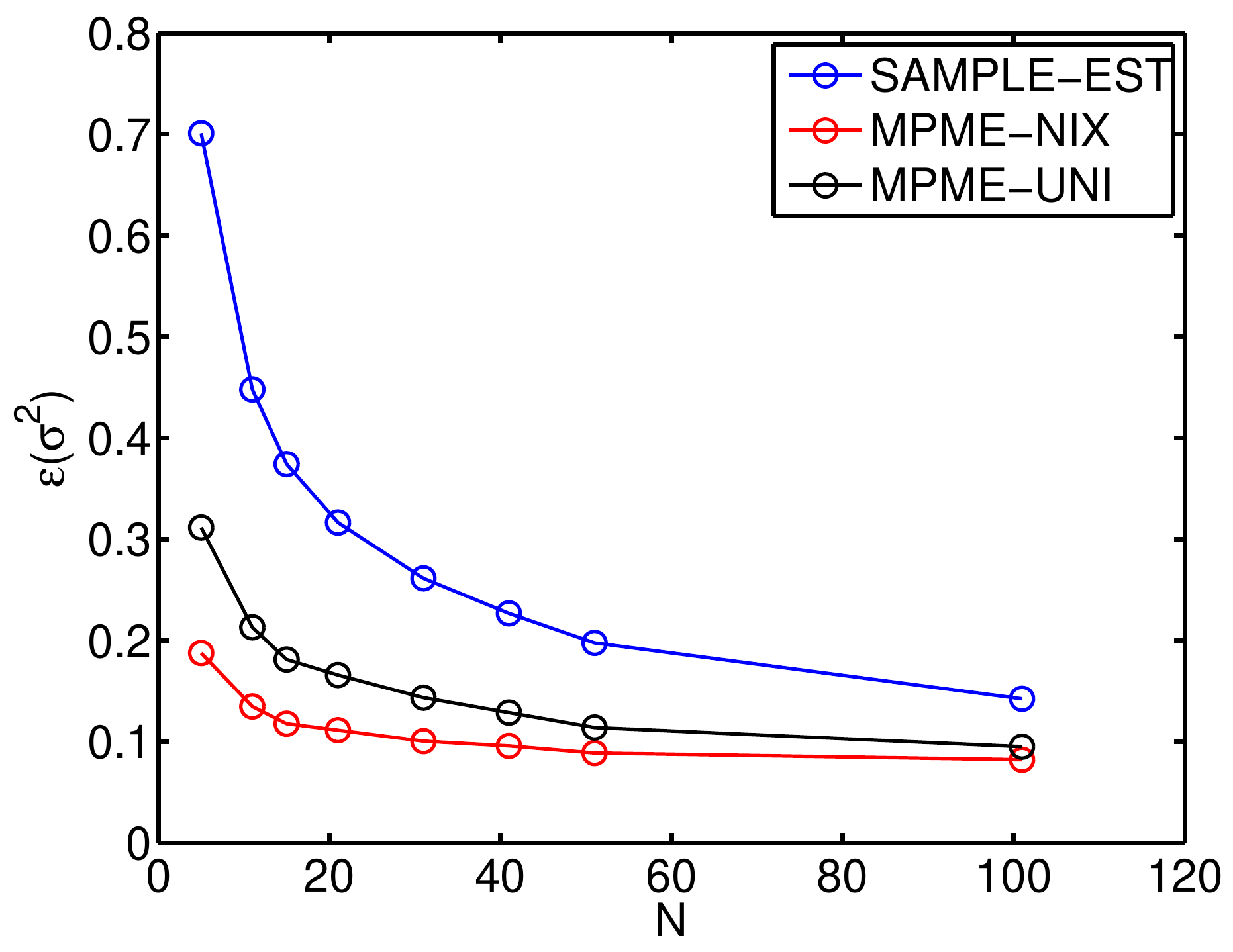}
                \caption{$\epsilon_{\sigma^2}$ vs $N$.}
        \end{subfigure}
  \caption{Comparison of sample estimators and MPME ($P=20$, Example 1).}
\label{fig:ex1case2}
\end{figure}

\figref{ex1case2} shows the error of three methods for different values of $N$ when $P=20$. It can be observed that as $N$ becomes large, the error of three methods all converge to a small value, and MPME does not present much advantage over the sample estimators. However, when $N$ is extremely small, MPME obtains significantly better accuracy.

\begin{figure}[h!]
  \centering
        \begin{subfigure}[b]{0.8\linewidth}
                \centering
    \includegraphics[width=\linewidth]{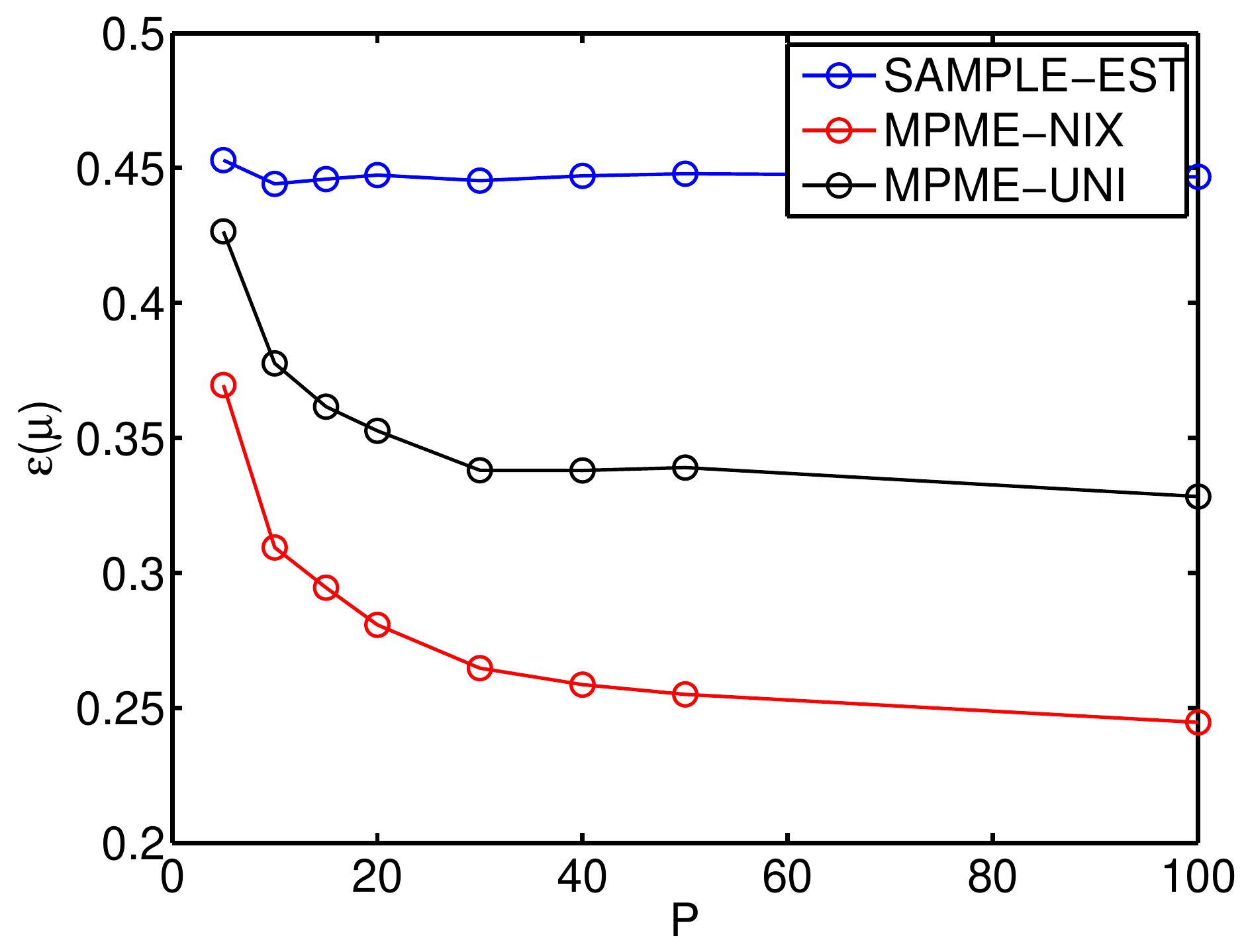}
                \caption{$\epsilon_{\mu}$ vs $P$.}
        \end{subfigure}%

        \begin{subfigure}[b]{0.8\linewidth}
                \centering
    \includegraphics[width=\linewidth]{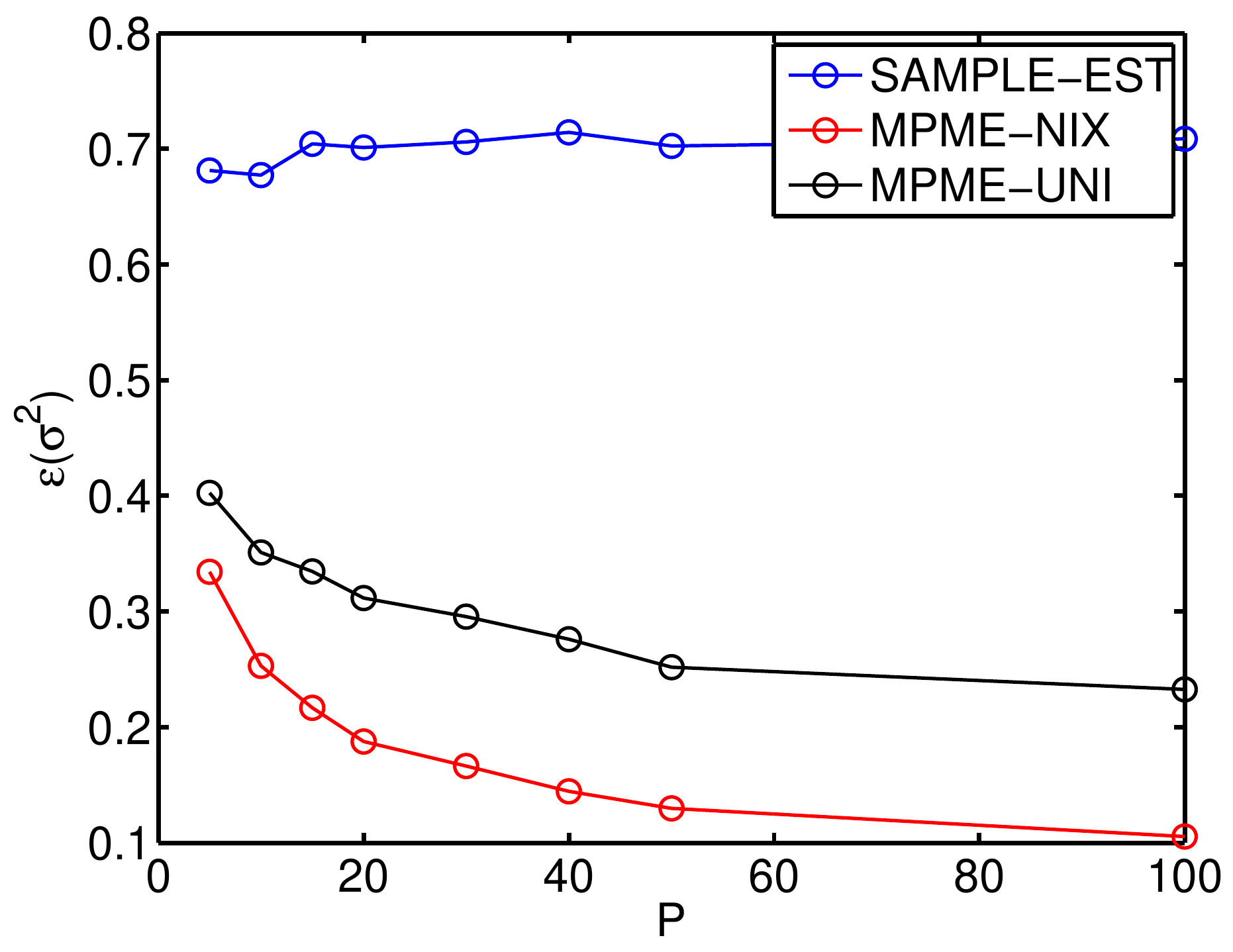}
                \caption{$\epsilon_{\sigma^2}$ vs $P$.}
        \end{subfigure}
  \caption{Comparison of sample estimators and MPME ($N=5$, Example 1).}
\label{fig:ex1case1}
\end{figure}

\figref{ex1case1} shows the error of three methods for different values of $P$ when $N=5$. It can be observed that as $P$ becomes large, the error of MPME decreases roughly as $1/\sqrt{P}$, while the error of the sample estimators stays the same.
The reason is that sample estimators treat each population independently, while MPME exploits the joint information in the dataset to improve the estimation accuracy at individual populations.

\begin{figure}[h!]
  \centering
        \begin{subfigure}[b]{0.8\linewidth}
                \centering
    \includegraphics[width=\linewidth]{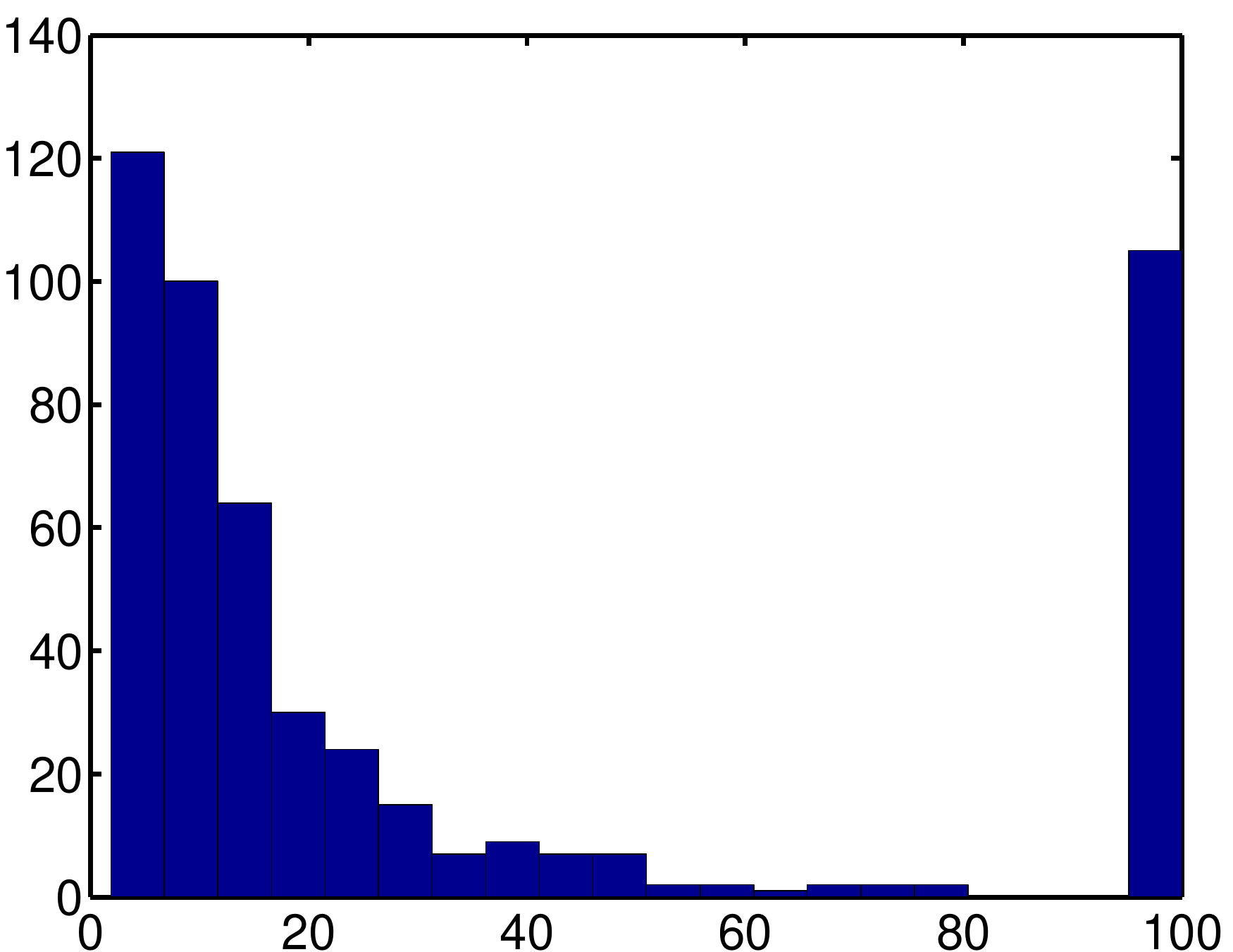}
                \caption{$\kappa_0$.}
        \end{subfigure}%

        \begin{subfigure}[b]{0.8\linewidth}
                \centering
    \includegraphics[width=\linewidth]{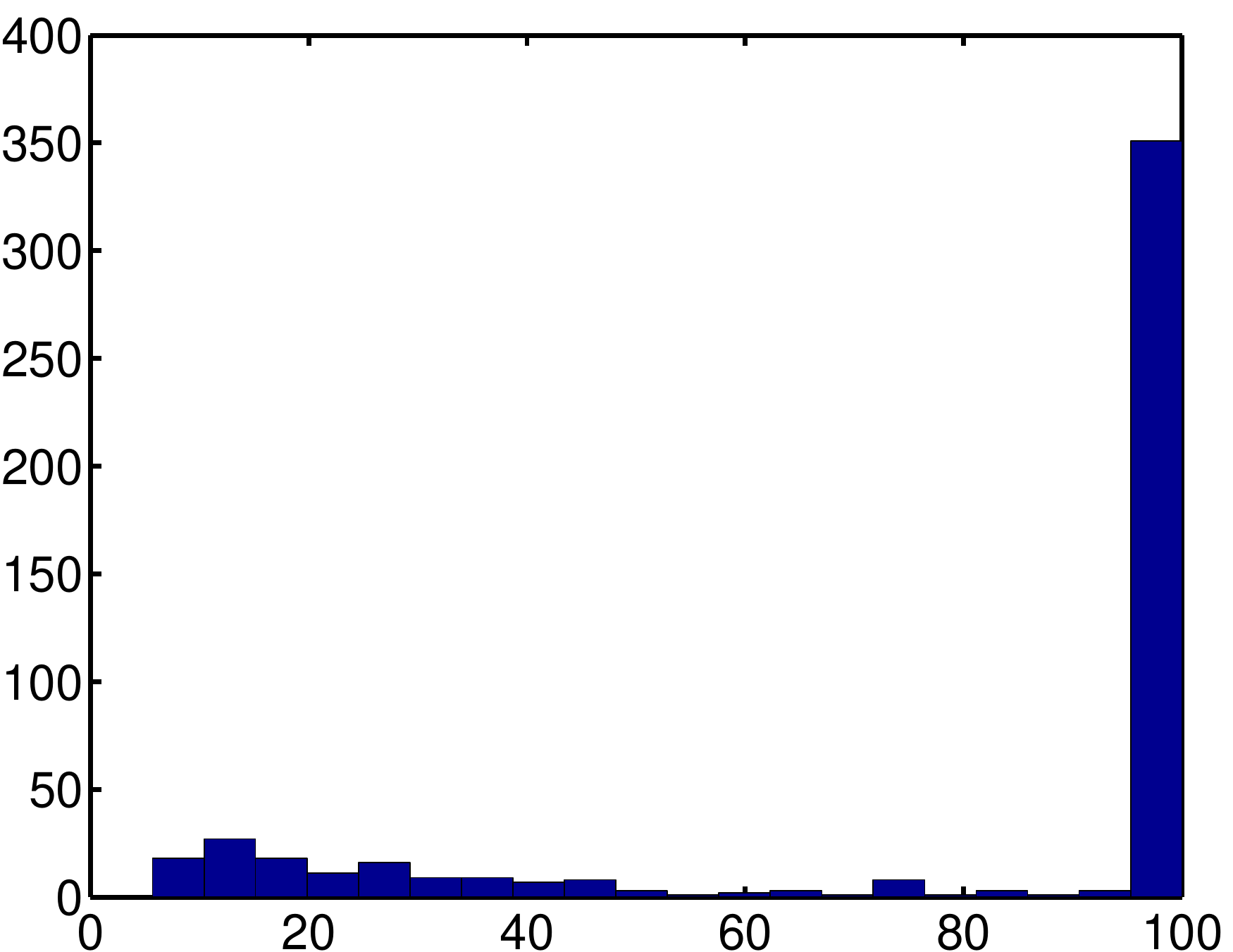}
                \caption{$\nu_0$.}
        \end{subfigure}
  \caption{Histogram of $\kappa_0$ and $\nu_0$, ($N=5, P=20$, Example 1).}
\label{fig:ex1hist1}
\end{figure}

As mentioned in \secref{secnix}, the application of the NIX prior can be interpreted as increasing the effective number of samples by $\kappa_0$ (for mean estimation) and $\nu_0$ (for variance estimation).
\figref{ex1hist1} shows the histogram of $\kappa_0$ and $\nu_0$ over 500 trials for the case $(N=5, P=20)$. The mean values of $\kappa_0$ and $\nu_0$ are 32.9 and 79.4, respectively.
This means, effectively, MPME increases the number of samples  $N=5$ to around 40 and 80 -- this significantly improves the accuracy of the estimation.

From \figref{ex1case2} and \figref{ex1case1}, we also find that for the MPME method, the NIX prior is usually better than the UNI prior, in terms of $\epsilon_{\mu}$ and $\epsilon_{\sigma^2}$. While this shows that NIX prior might be preferred, we emphasize that the UNI prior could lead to a better accuracy for a particular population.

\begin{figure}[h!]
  \centering
        \begin{subfigure}[b]{0.8\linewidth}
                \centering
    \includegraphics[width=\linewidth]{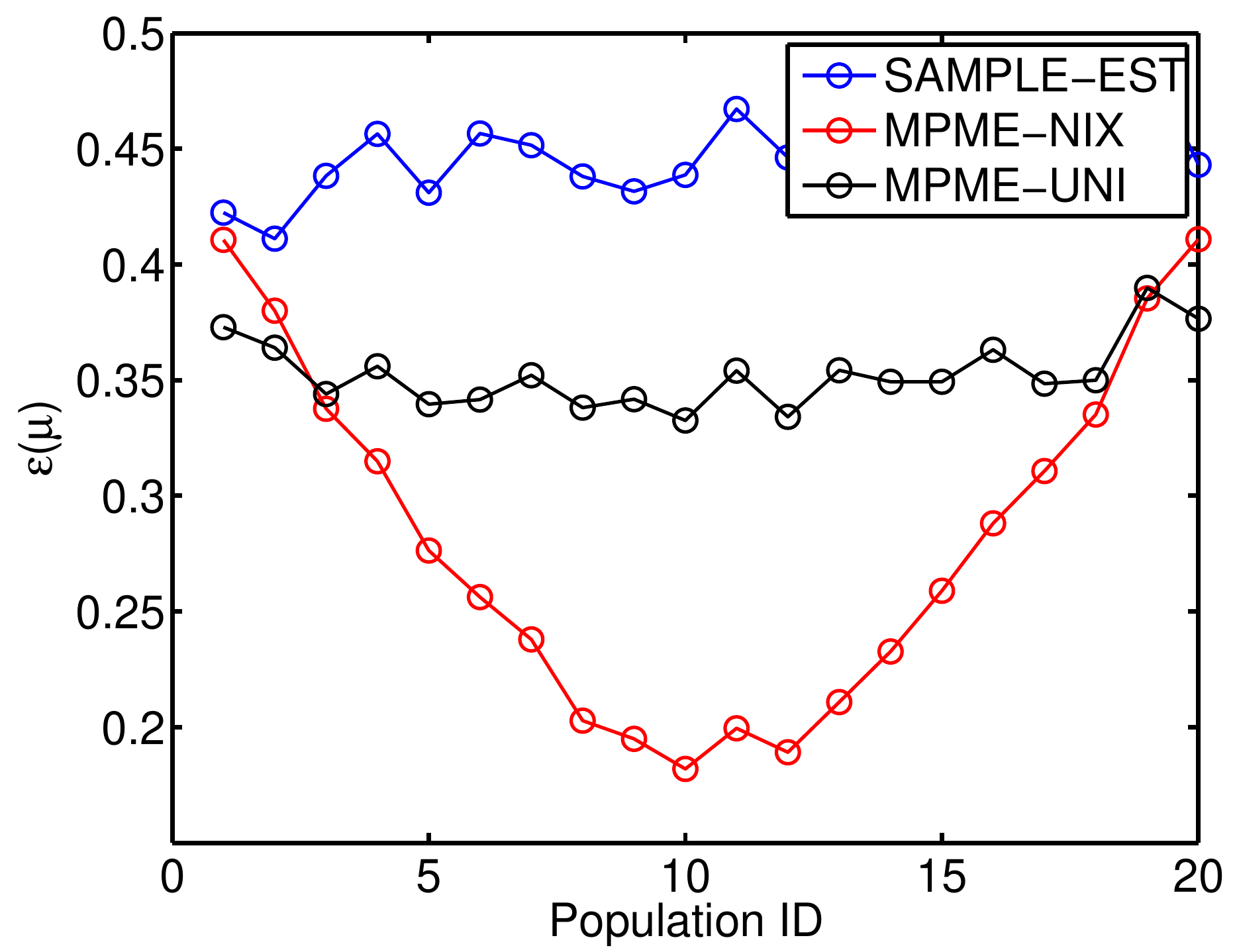}
                \caption{$\epsilon_{\mu}$ across populations.}
        \end{subfigure}%

        \begin{subfigure}[b]{0.8\linewidth}
                \centering
    \includegraphics[width=\linewidth]{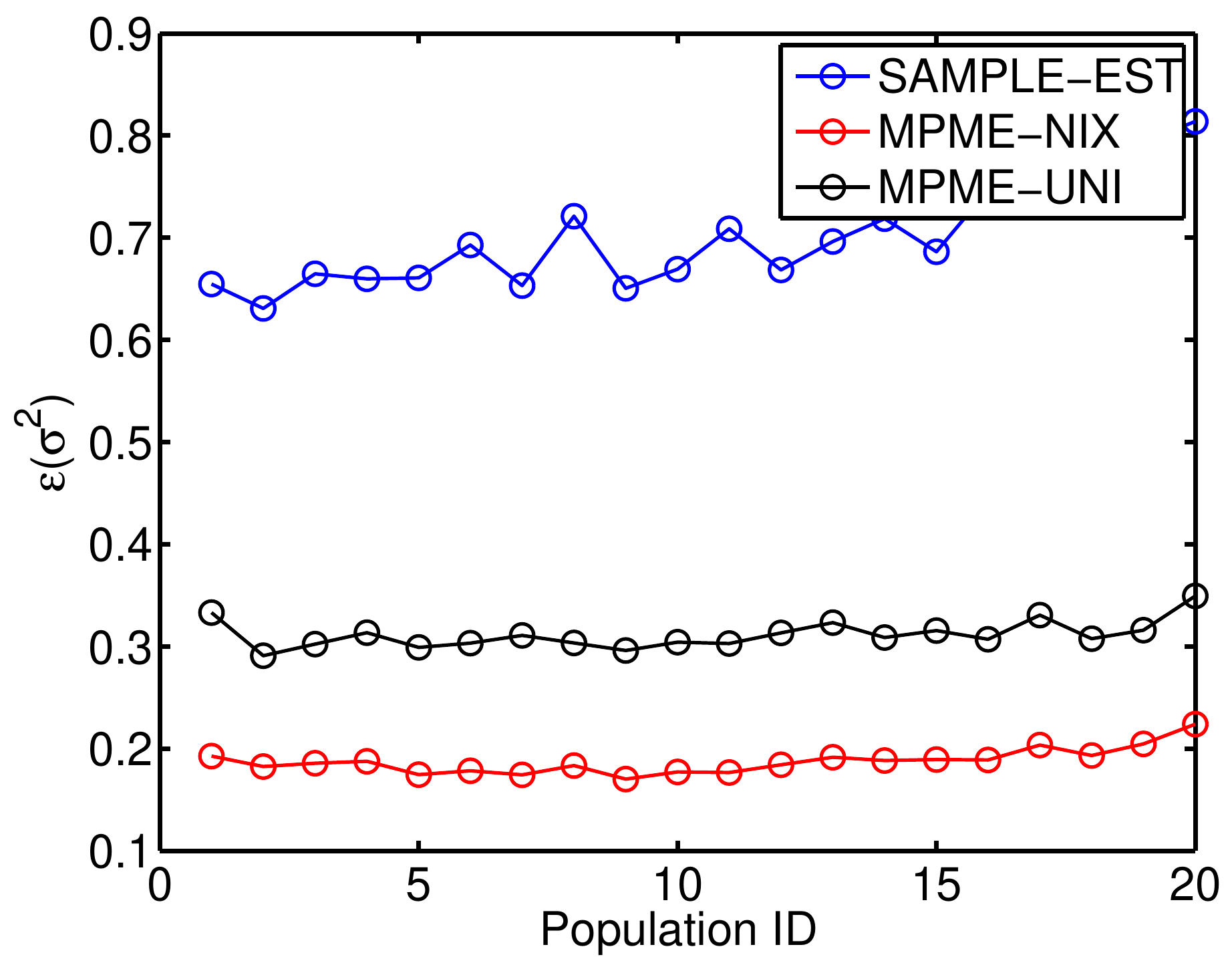}
                \caption{$\epsilon_{\sigma^2}$ across populations.}
        \end{subfigure}
  \caption{Comparison of sample estimators and MPME ($N=5, P=20$, Example 1).}
\label{fig:ex1individual}
\end{figure}

For example, \figref{ex1individual} shows the average error for each of the 20 populations for the setting $(N=5, P=20)$. 
For the estimation of $\sigma_i^2$'s, MPME-NIX is consistently better than MPME-UNI. However, for the estimation of $\mu_i$'s, although
the NIX prior leads to a smaller error for most of the populations, the UNI prior does better at the populations that have extreme $\mu_i$ values.
Intuitively, during the second step (MAP) in MPME, the UNI prior applies lower/upper bounds on the estimated mean, and the NIX prior pulls the estimated mean towards the joint mean (across populations). Therefore, if the population mean is close to the overall mean, NIX prior leads to a better estimation. On the other hand, if the population mean is far from the overall mean (\eg, at extreme corners), UNI prior will be better. 
In both cases, however, MPME-UNI and MPME-NIX are always better than the sample estimators.

\subsection{Synthetic Examples 2}
In this example, we use almost the same setting as the previous one, except for $\sigma_i$ values for the $i$-th population.
We choose $\sigma_i$ to be equally spaced over $[1.9, 2.1]$, \ie, $\sigma_i = 1.9 + \frac{0.2}{P-1} (i-1)$, \ie, twice of $\sigma_i$ in the previous example.

\begin{figure}[h!]
  \centering
        \begin{subfigure}[b]{0.8\linewidth}
                \centering
    \includegraphics[width=\linewidth]{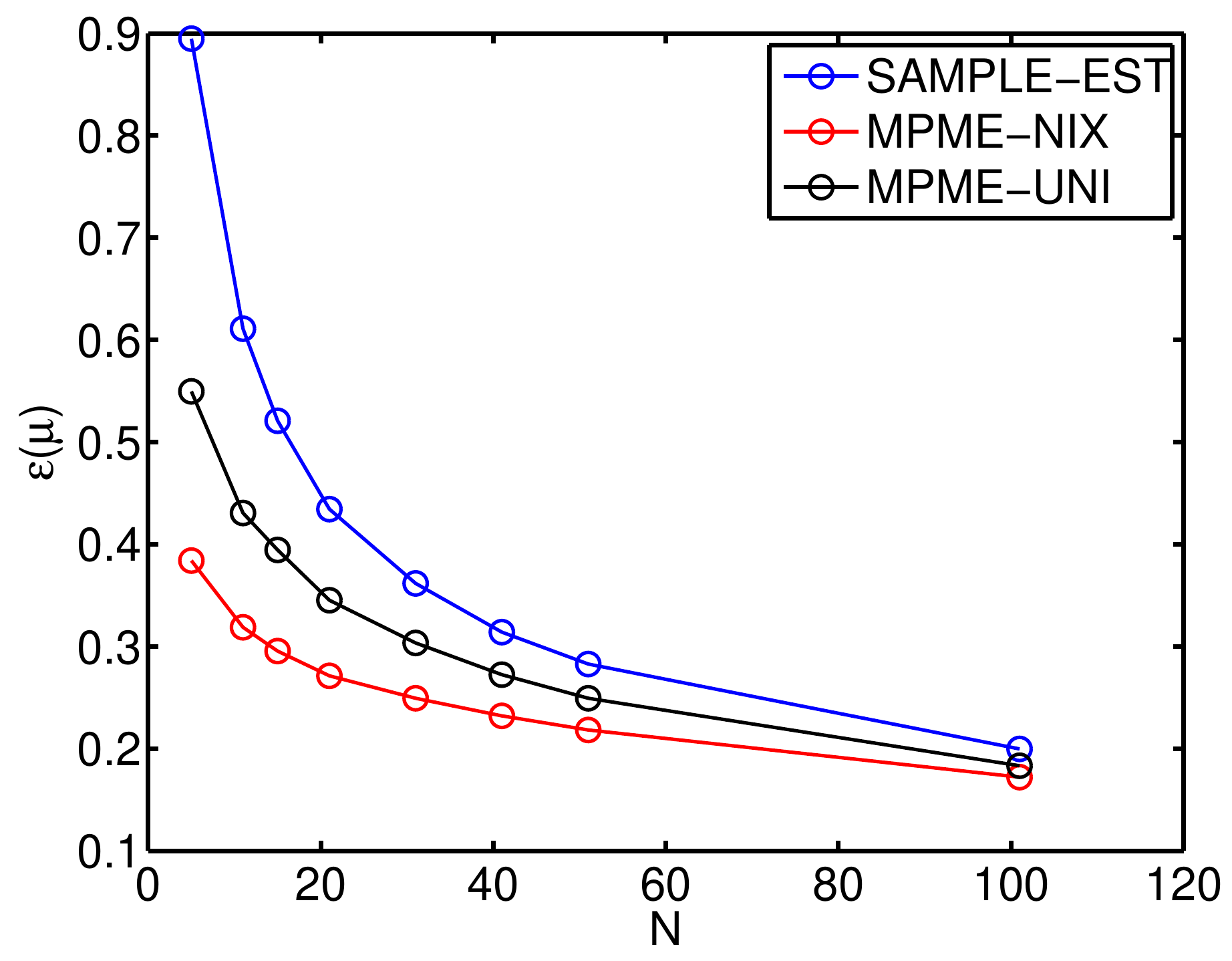}
                \caption{$\epsilon_{\mu}$ vs $N$.}
        \end{subfigure}%

        \begin{subfigure}[b]{0.8\linewidth}
                \centering
    \includegraphics[width=\linewidth]{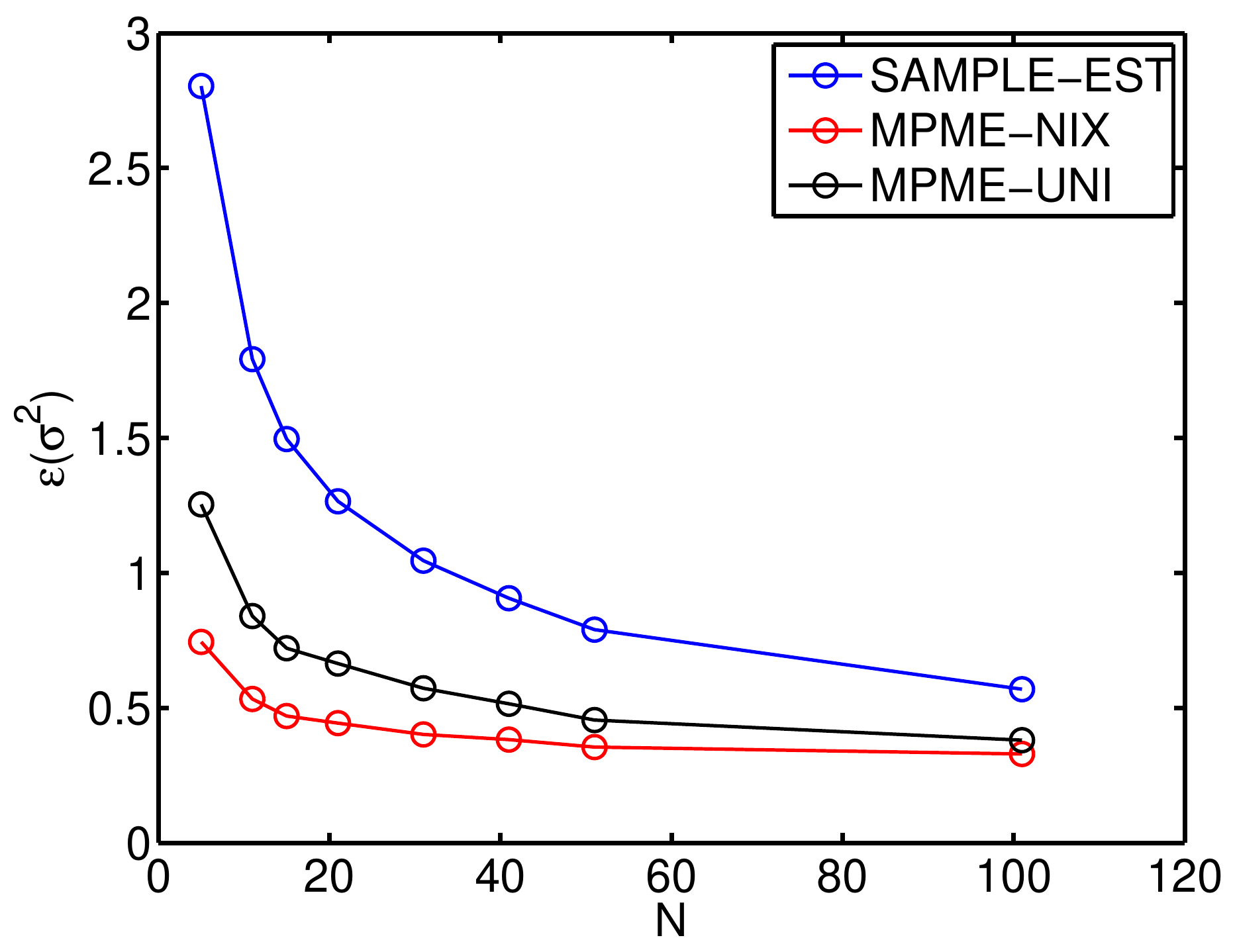}
                \caption{$\epsilon_{\sigma^2}$ vs $N$.}
        \end{subfigure}
  \caption{Comparison of sample estimators and MPME ($P=20$, Example 2).}
\label{fig:ex2case2}
\end{figure}

\begin{figure}[h!]
  \centering
        \begin{subfigure}[b]{0.8\linewidth}
                \centering
    \includegraphics[width=\linewidth]{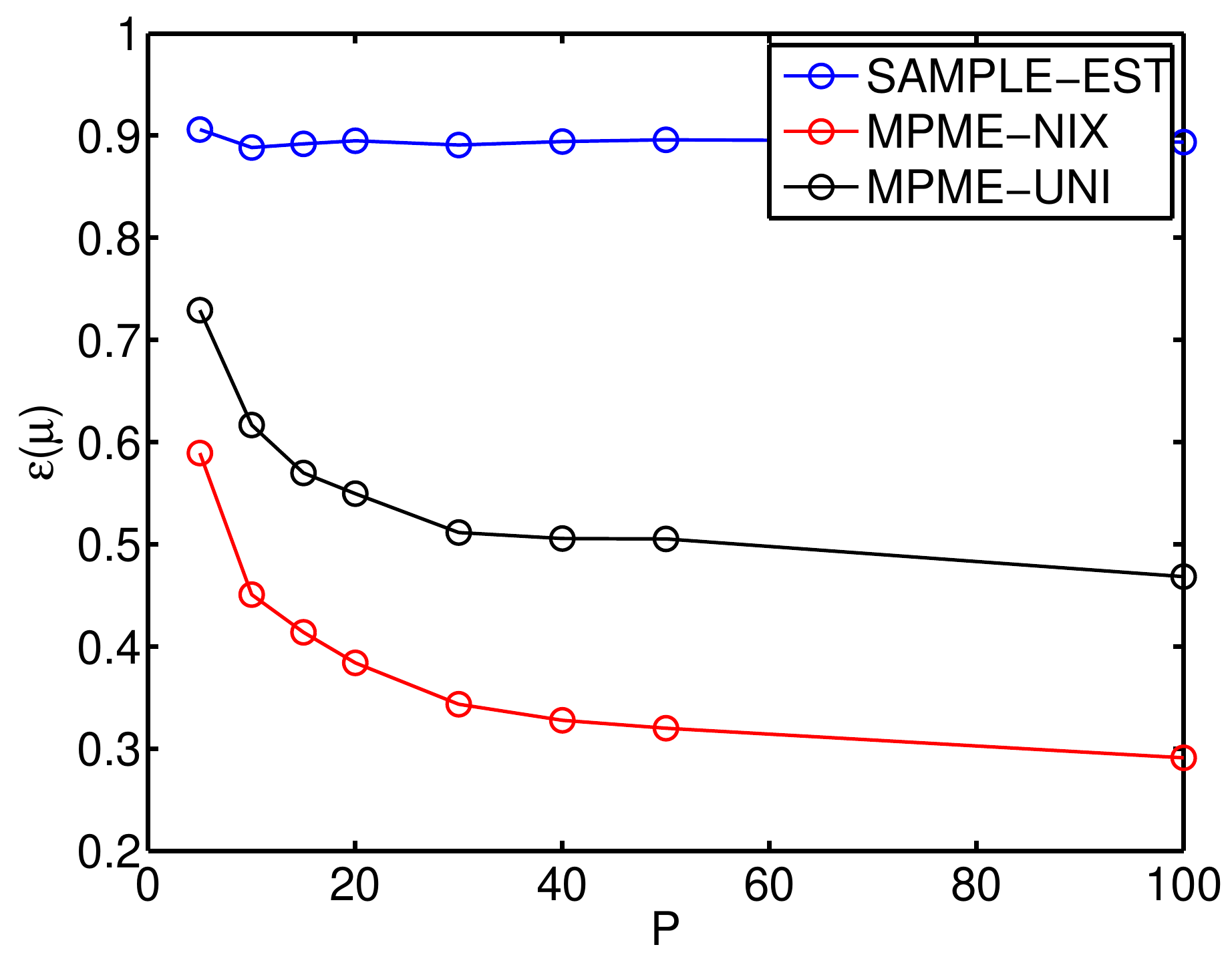}
                \caption{$\epsilon_{\mu}$ vs $P$.}
        \end{subfigure}%

        \begin{subfigure}[b]{0.8\linewidth}
                \centering
    \includegraphics[width=\linewidth]{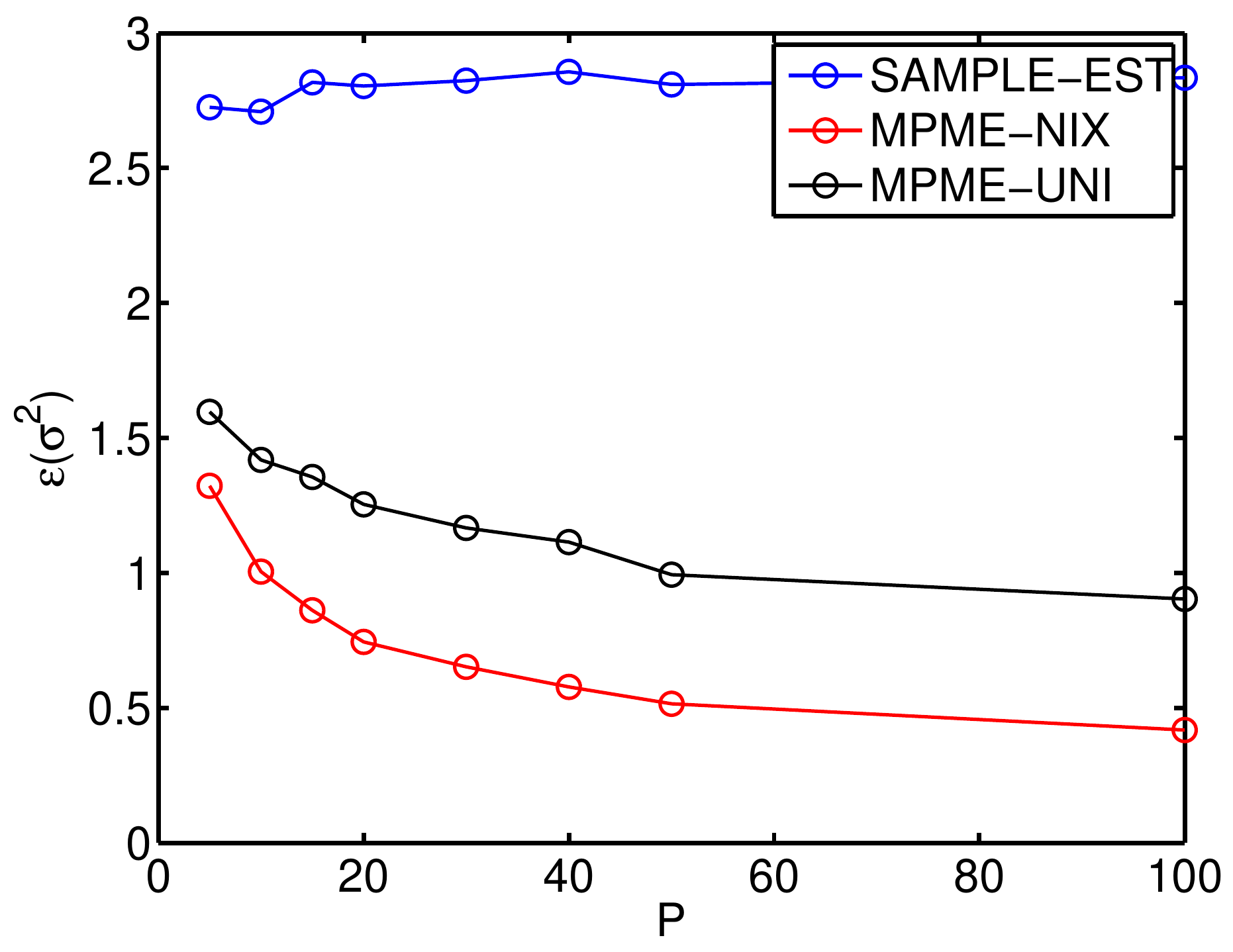}
                \caption{$\epsilon_{\sigma^2}$ vs $P$.}
        \end{subfigure}
  \caption{Comparison of sample estimators and MPME ($N=5$, Example 2).}
\label{fig:ex2case1}
\end{figure}

Similar trends are observed as in the previous example, as shown in \figref{ex2case2} and \figref{ex2case1}. However, compared to the previous example, it is worthwhile to note that MPME obtains relatively more reduction in error over sample estimators.
The reason for this is that when the variance at each population is smaller, the data show less ``uncertainty/randomness''. For example, the sample mean estimator has a confidence interval proportional to $\sigma_i$ and the sample variance estimator has a confidence interval proportional to $\sigma_i^2$.
Therefore, when $\sigma_i$'s are small, sample estimators achieve relatively good accuracy, and MPME provides less improvement.
However, when $\sigma_i$'s are large, MPME beats sample estimators significantly, by exploiting the collective information gathered from multiple populations.

\subsection{Validation of a High-Speed I/O Link}
\label{sec:ioexample}

In I/O link validation, one critical performance metric is Bit-Error-Ratio (BER). For the state-of-the-art high-speed links, the BER is extremely small. For example, in the latest PCIE  specification \cite{pcisig},
 $\text{BER}_{\text{spec}} = 10^{-12}$ with 8Gb$/$sec data rate. This makes BER measurement a very time-consuming process. An alternative is to measure the eye width and eye height (\aka, \textit{time margin} (TM) and \textit{voltage margin} (VM), respectively) of the eye diagram at the receiver, which can be converted to BER under reasonable assumptions. Margin measurement, although much faster than direct BER measurement, is still expensive in terms of time and cost. For a limited time period, only a small number of data can be measured for each configuration.

\begin{figure}[h!]
  \centering
    \includegraphics[width=0.8\linewidth]{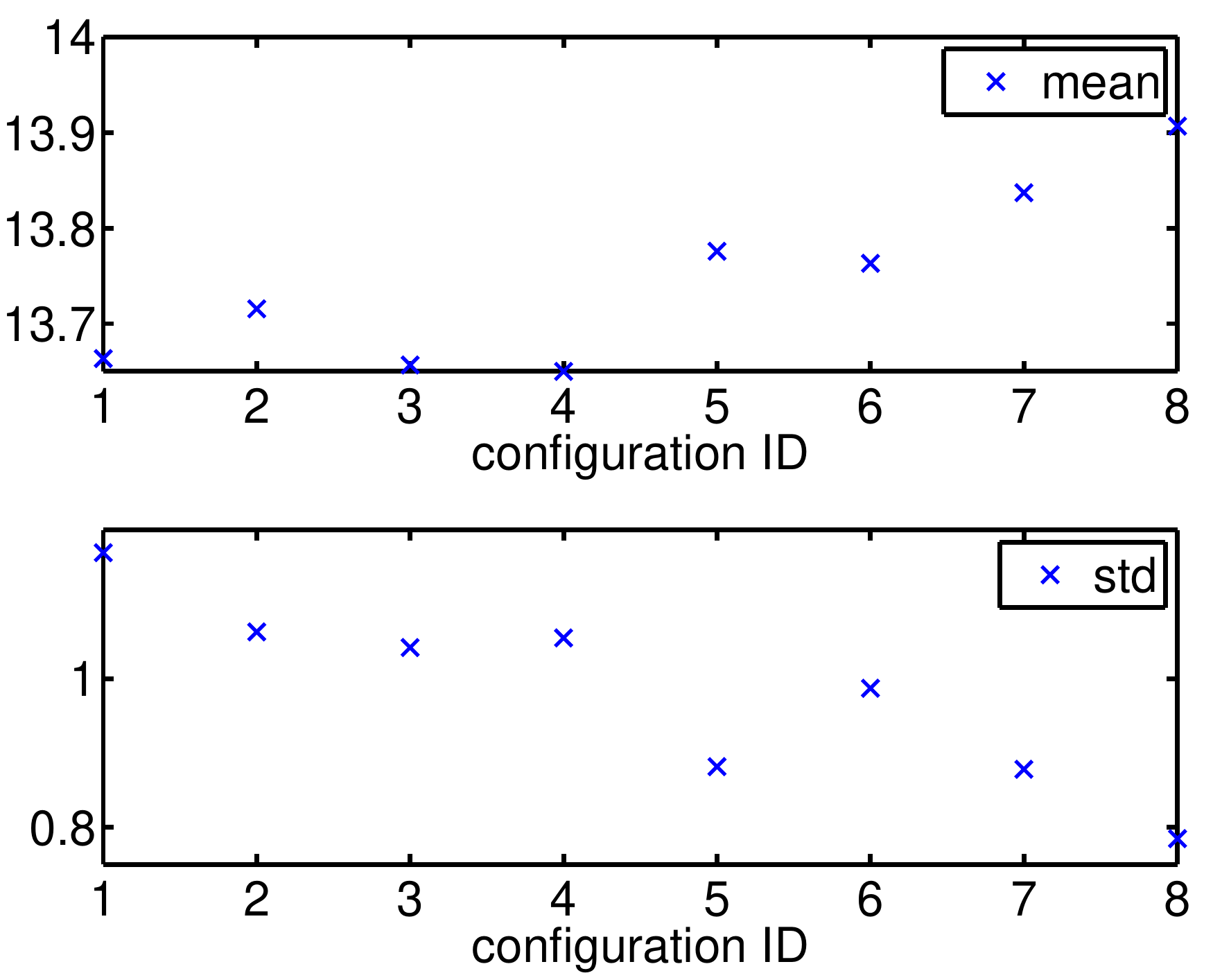}
  \caption{Mean and standard deviation at 8 configurations.}
\label{fig:ex3stat}
\end{figure}

In this example, we have measured the time margin of 50 dies (randomly sampled) for 8 different configurations.
(Note that we measured 50 dies simply for the purpose of validating our algorithm. In practice, only about 5 dies might be measured.)
The mean and standard deviation at different configurations are shown in \figref{ex3stat}.
We have also observed from the histogram that the distribution of time margin can be well approximated by Gaussian distributions.

To compare the results of MPME and sample estimators, we take $N$ samples of data for each configuration from the 50 measurements, and apply both methods. We repeat this experiment for 500 times, and compare the statistics of $\epsilon_{\mu}$ and $\epsilon_{\sigma^2}$. 
\revA{This is also known as bootstrap in statistics literature \cite{bishop2006pattern}.}

\begin{figure}[h!]
  \centering
        \begin{subfigure}[b]{0.8\linewidth}
                \centering
    \includegraphics[width=\linewidth]{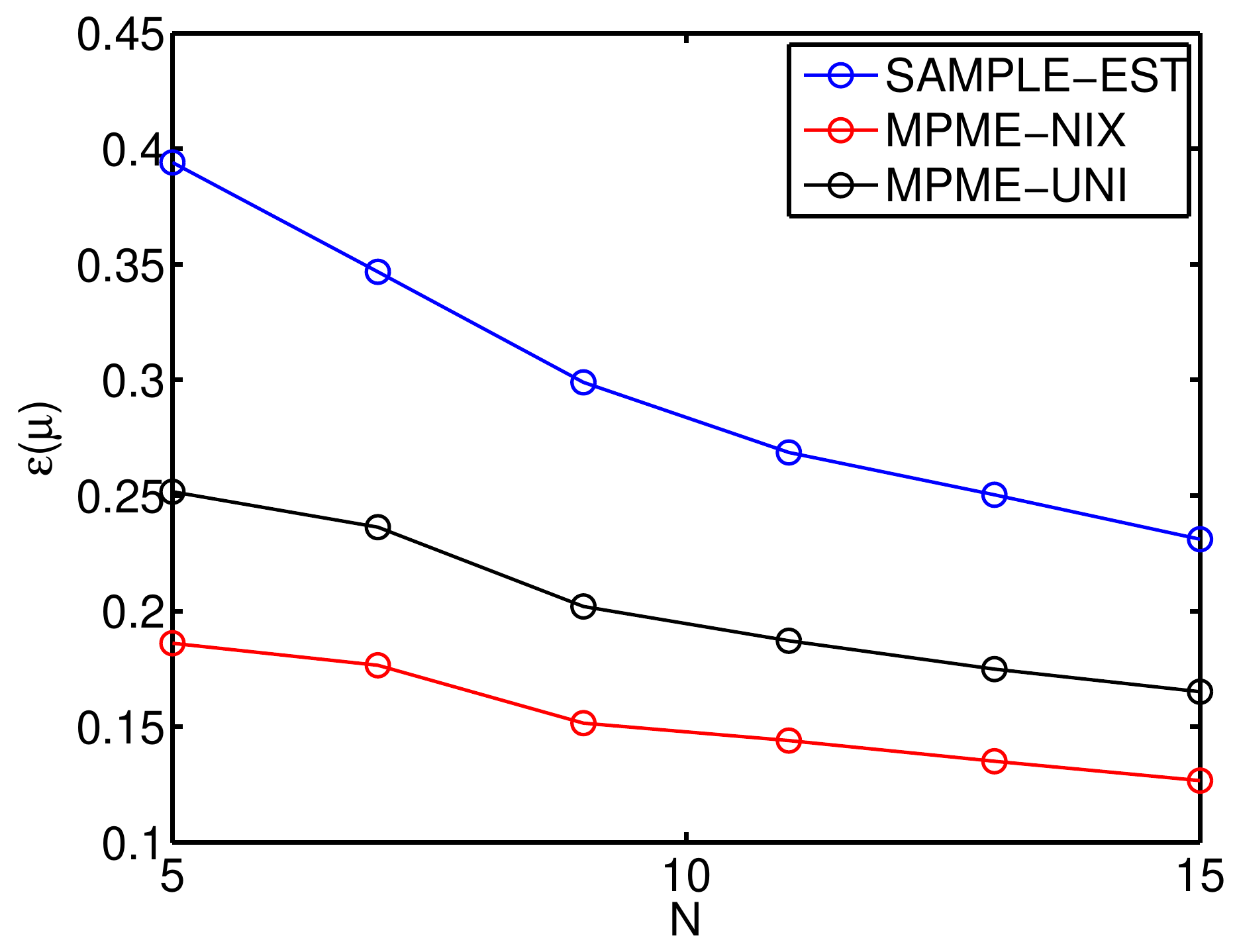}
                \caption{$\epsilon_{\mu}$ vs $N$.}
        \end{subfigure}%

        \begin{subfigure}[b]{0.8\linewidth}
                \centering
    \includegraphics[width=\linewidth]{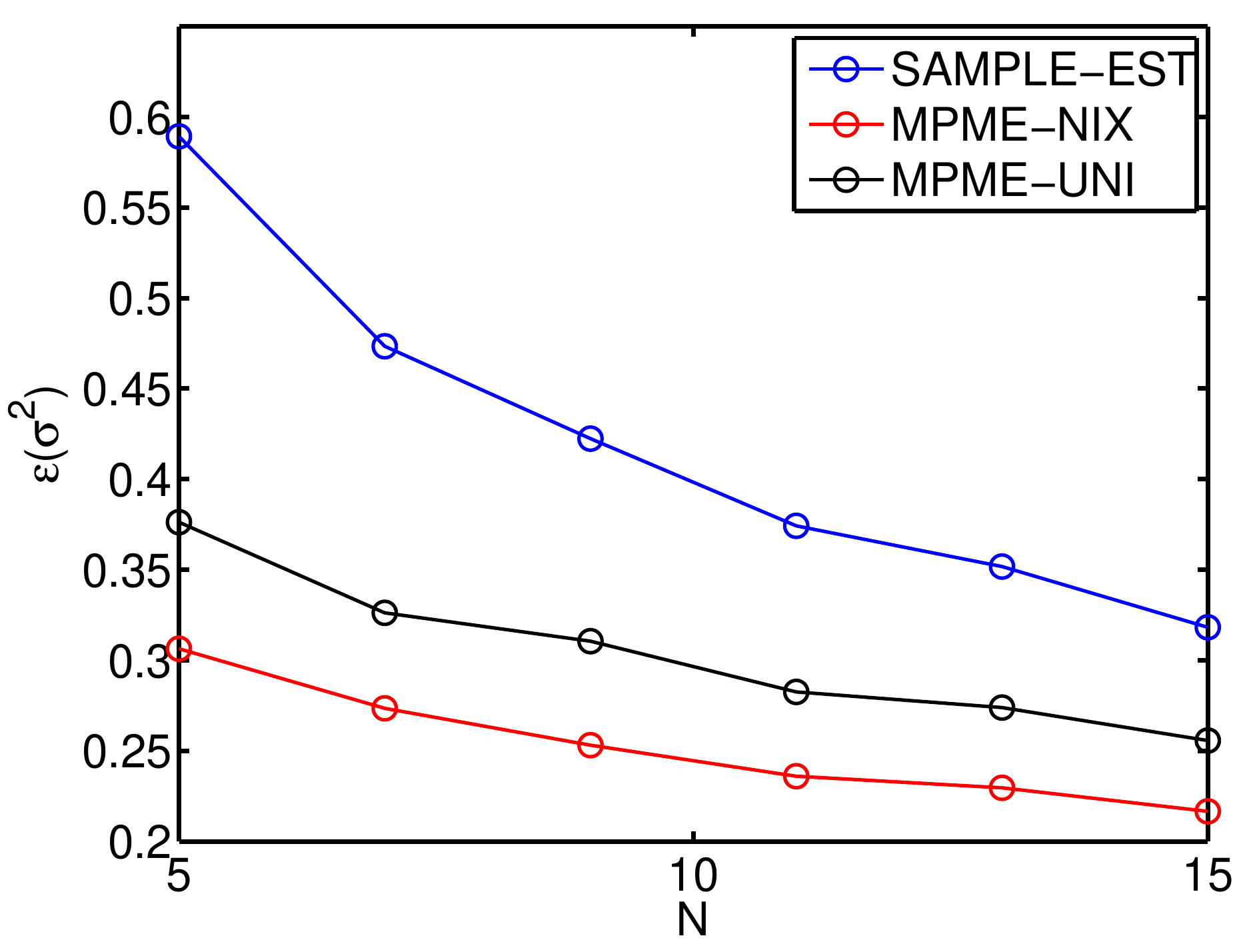}
                \caption{$\epsilon_{\sigma^2}$ vs $N$.}
        \end{subfigure}
  \caption{Comparison of sample estimators and MPME.}
\label{fig:intelex1}
\end{figure}

The results for $\epsilon_{\mu}$ and $\epsilon_{\sigma^2}$ for different values of $N$ are plotted in \figref{intelex1}. 
Similar to the synthetic examples, it is observed that when the sample size is small, sample estimators are much less accurate than MPME, thus may lead to unreliable validation conclusions.

\ignore{
\begin{table*}[th]
\centering
\caption{Experimental results of $\epsilon_{\mu}$ and $\epsilon_{\sigma^2}$ of the high-speed I/O example.}
\begin{tabular}{c|c|c|c|c|c|c}
\hline
N & SAMPLE-EST($\mu$) & MPME-NIX($\mu$) & MPME-UNI($\mu$) & SAMPLE-EST($\sigma^2$) & MPME-NIX($\sigma^2$) & MPME-UNI($\sigma^2$) \\
\hline
5 & 0.2722 & 0.0675 & 0.1276 & 0.1806 & 0.07753 & 0.1142 \\
7 & 0.2113 & 0.05784 & 0.11 & 0.1281 & 0.06527 & 0.08789 \\
9 & 0.1611 & 0.04538 & 0.08187 & 0.1046 & 0.05783 & 0.07687 \\
11 & 0.1264 & 0.0396 & 0.07015 & 0.08864 & 0.05432 & 0.06906 \\
13 & 0.1113 & 0.03364 & 0.05878 & 0.082 & 0.05386 & 0.06672 \\
15 & 0.09377 & 0.02929 & 0.0514 & 0.06872 & 0.04999 & 0.05936 \\
\hline
\end{tabular}
\label{tab:intelex1}
\end{table*}
}

Besides accuracy improvement, \figref{intelex1} shows another practical implication of MPME -- for the same overall accuracy, MPME requires much less samples than the sample estimators.  In this particular example, MPME-NIX would need just about 50\% samples than the sample estimators, in order to obtain the same accuracy. It directly implies less validation time, and thus faster product time-to-market.

\begin{figure}[h!]
  \centering
        \begin{subfigure}[b]{0.8\linewidth}
                \centering
    \includegraphics[width=\linewidth]{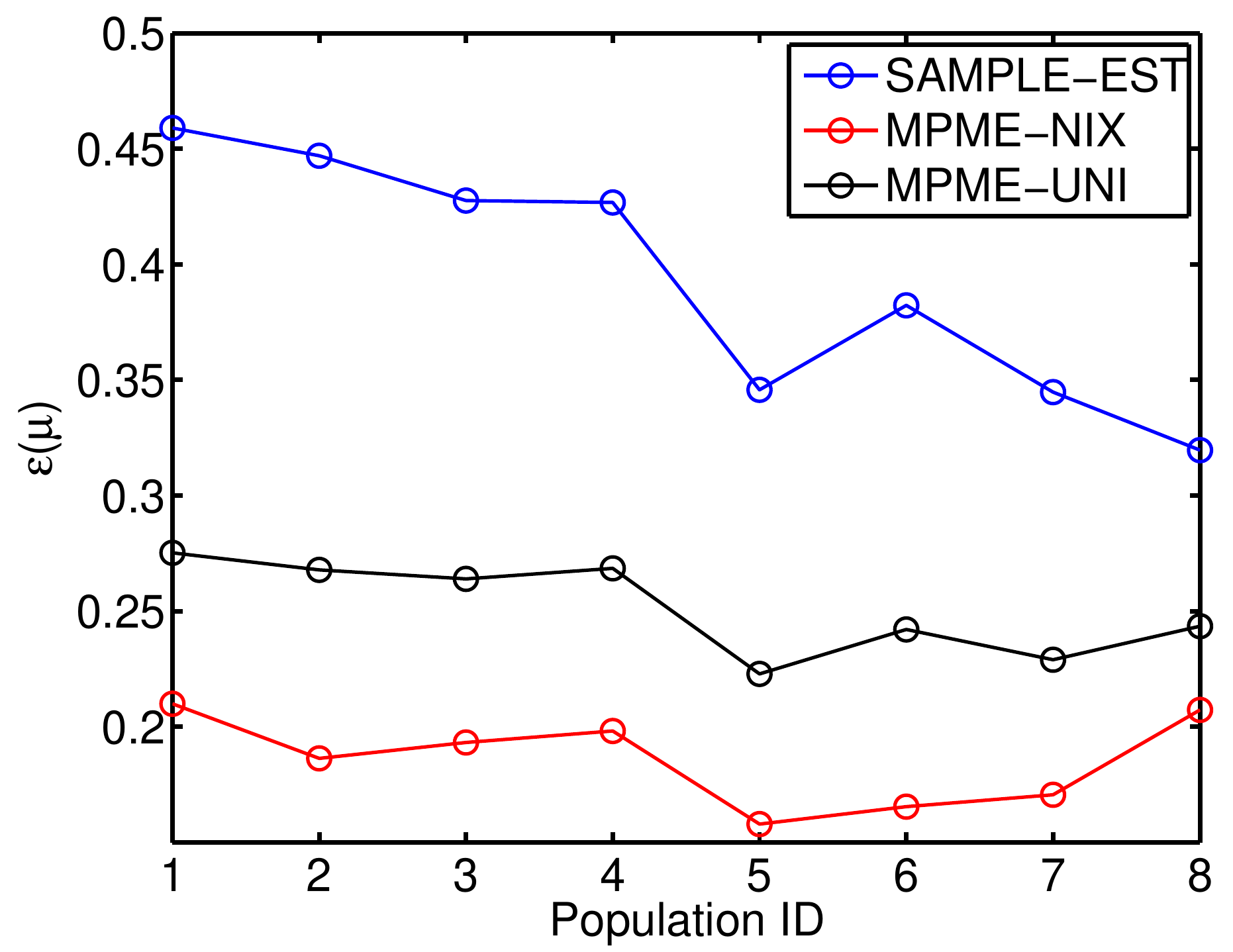}
                \caption{$\epsilon_{\mu}$ across populations.}
        \end{subfigure}%

        \begin{subfigure}[b]{0.8\linewidth}
                \centering
    \includegraphics[width=\linewidth]{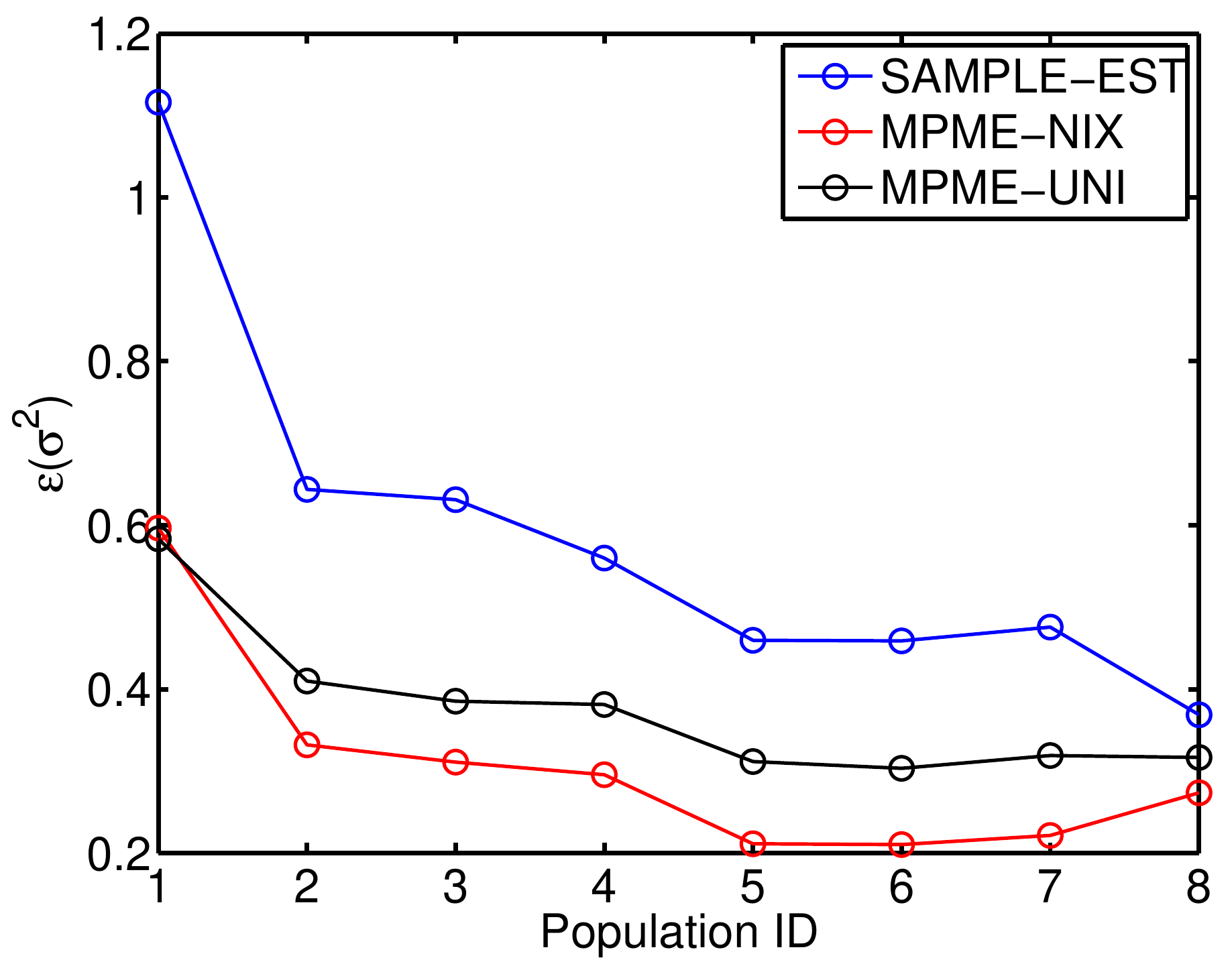}
                \caption{$\epsilon_{\sigma^2}$ across populations.}
        \end{subfigure}
  \caption{Comparison of sample estimators and MPME.}
\label{fig:intelindividual}
\end{figure}

\figref{intelindividual} shows a detailed comparison of $\epsilon_{\mu}$ and $\epsilon_{\sigma^2}$ for all 8 populations, and the results confirm our conclusion drawn from synthetic examples -- \ie, if the variance of the population is larger, MPME is more effective in reducing the error. In particular, the population \#1  in this example has the largest variance, and MPME has the most significant error reduction over the sample estimators.

\section{Conclusion}
In this paper, we have proposed MPME, an efficient method for estimating moments (mean and variance) of multiple populations.
The key difficulty we try to address is the problem of extremely small sample size, which is commonly seen in circuit validation.
MPME alleviates this problem by considering samples obtained from many populations, which in practice can refer to different corners and configurations.
MPME leverages data from all populations to improve the estimation accuracy for each population, and the method fits nicely under the hierarchical Bayesian framework.
We validated MPME on several datasets, including measurement of a commercial I/O link.
We show that MPME is consistently better than the sample mean/variance estimators, and can achieve up to 2$\times$ average accuracy improvement. Furthermore, the accuracy improvement can also be equivalently translated to a potentially large test/validation time reduction.

\appendices

\section{Probabilistic Graphical Models}
\seclabel{graphicalmodels}

Probabilistic graphical models use graphs  (directed or undirected) to describe multi-variate probability distributions and the probabilistic structures (\eg, conditional independences).
For the interest of this paper, we only discuss the concepts and notations relevant to the MPME method.
For more details about graphical models, we refer the readers to two excellent books \cite{bishop2006pattern, kollar2009probabilistic}.

\begin{figure}[h!]
  \centering
        \begin{subfigure}[b]{0.49\linewidth}
                \centering
    \includegraphics[width=\linewidth]{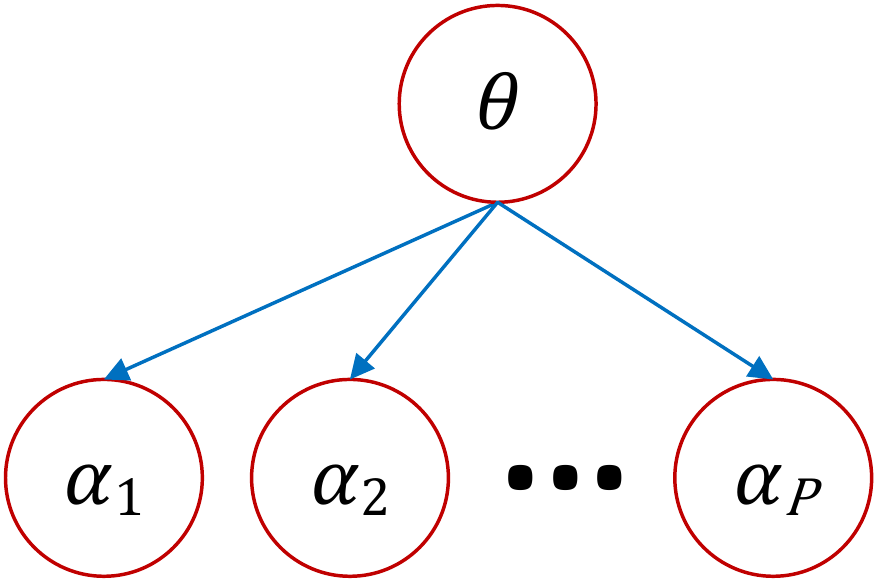}
                \caption{A simple directed graphical model.}
\label{fig:introgm1}
        \end{subfigure}%
\quad
        \begin{subfigure}[b]{0.23\linewidth}
                \centering
    \includegraphics[width=\linewidth]{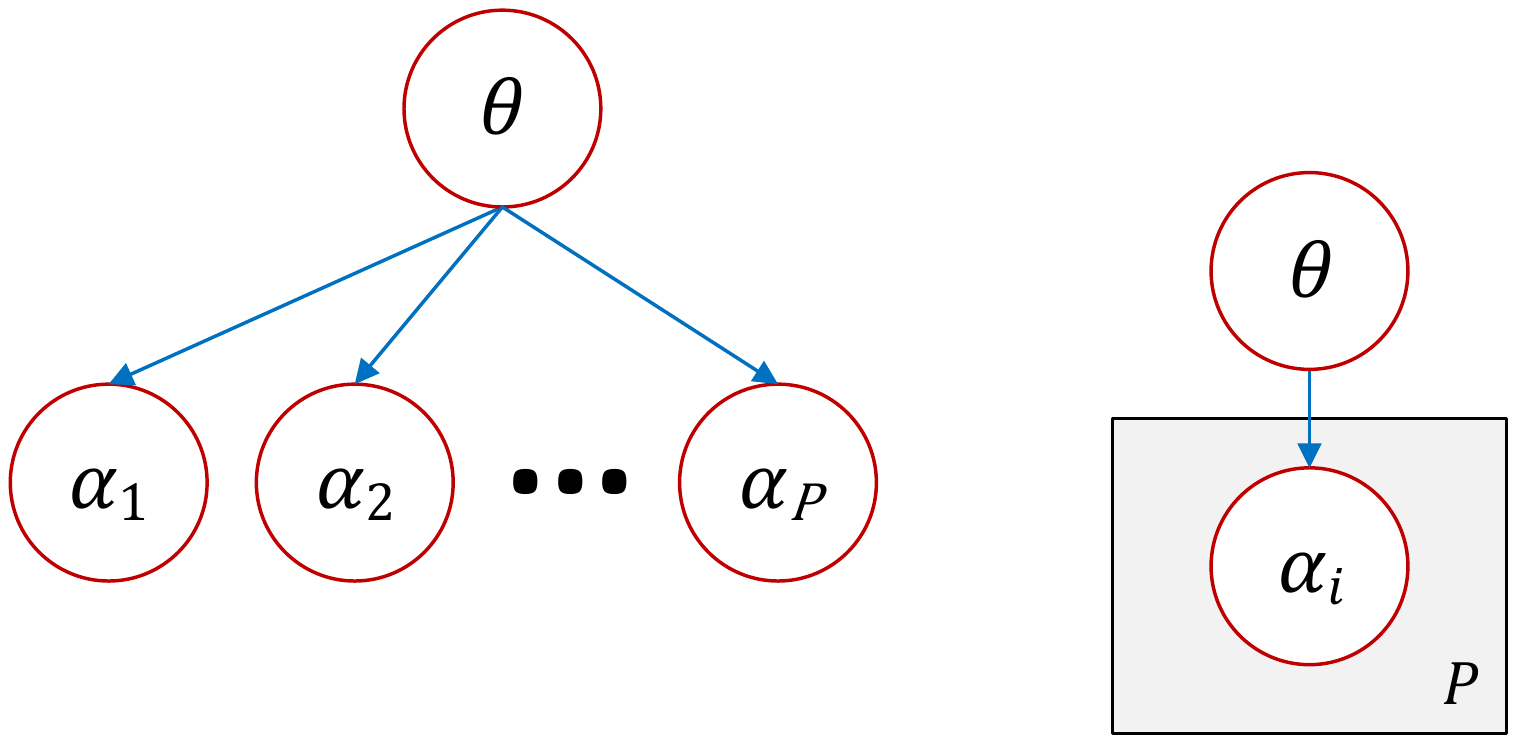}
                \caption{Plate notation.}
\label{fig:introgm2}
        \end{subfigure}
  \caption{A simple tree-like graphical model.}
\label{fig:introgm}
\end{figure}

In a graphical model, each node represents a random variable (or a set of random variables), and the edges represent the probabilistic relationships between these variables.
In a directed graphical model, the edges can be interpreted as the dependency among variables. For example, a tree-like graphical model shown in \figref{introgm1} describes a joint probability distribution over $\theta, \alpha_1, \cdots, \alpha_P$ as
\begin{equation}
p(\theta, \alpha_1, \cdots, \alpha_P) = p(\theta) p(\alpha_1|\theta) \cdots p(\alpha_P|\theta),
\end{equation}
which encodes the conditional independence
\begin{equation}
\alpha_1 \ci \alpha_2 \ci \cdots \ci \alpha_P | \theta.
\end{equation}
Here, the notation $(A\ci B|C)$ means that $A$ and $B$ are conditionally independent given $C$.

To simplify the graph notation, we use the \textit{plate notation} to compactly represent multiple nodes.
In the plate notation, we draw a single representative node and then surround it with a box labeled with $P$ indicating there are $P$ nodes of this kind.
Using the plate notation, the graphical model in \figref{introgm1} can be compactly represented by \figref{introgm2}.

\section{Correlation Induced by Imposing a Prior Distribution}
\label{sec:priorcorr}

In this section, we explain why random variables that are conditionally independent are correlated, and how that relates to the traditional concept of correlation coefficient.

For simplicity, we consider the graphical model in \figref{introgm1} where there are only two leaf nodes $\alpha_1$ and $\alpha_2$. We further assume that
\begin{equation}
\begin{aligned}
\begin{bmatrix}
\alpha_1\\
\alpha_2 \\
\end{bmatrix} \bigg| \theta 
\sim  \mc{N}\left(
\begin{bmatrix}
1\\
1\\
\end{bmatrix} \theta, 
\begin{bmatrix}
\sigma^2 & 0 \\
0 & \sigma^2 \\
\end{bmatrix}
\right),
\end{aligned}
\label{eqn:co1}
\end{equation}
where $\sigma$ is known. \eqnref{co1} implies that $\alpha_1$ and $\alpha_2$ are conditionally independent given $\theta$.

However, since $\theta$ is not observed, we need to study the marginal distribution of $(\alpha_1, \alpha_2)$ to compute their correlation coefficient.
To compute the marginal distribution, we assume that $\theta$ also follows a Gaussian distribution
\begin{equation}
\theta \sim \mc{N}(\mu_0, \sigma_0^2).
\end{equation}

Then, we can compute 
\begin{equation}
\begin{aligned}
p(\alpha_1, \alpha_2) =& \int_{\theta} p(\alpha_1, \alpha_2, \theta) d\theta \\
=& \int_{\theta} p(\alpha_1, \alpha_2 | \theta) p(\theta) d\theta .
\end{aligned}
\end{equation}
With some algebraic manipulation, we obtain
\begin{equation}
\begin{bmatrix}
\alpha_1\\
\alpha_2\\
\end{bmatrix} \sim
\mc{N}\left(
\begin{bmatrix}
1\\
1\\
\end{bmatrix} \mu_0, 
\begin{bmatrix}
\sigma^2 & 0 \\
0 & \sigma^2 \\
\end{bmatrix}
+ \sigma_0^2\begin{bmatrix}
1 & 1\\
1 & 1\\
\end{bmatrix} 
\right).
\end{equation}

Therefore, the correlation coefficient between $\alpha_1$ and $\alpha_2$ is
\begin{equation}
\rho = \frac{\sigma_0^2}{\sigma^2 + \sigma_0^2} = \frac{1}{1+\frac{\sigma^2}{\sigma_0^2}}.
\label{eqn:co2}
\end{equation}

From \eqnref{co2}, we can conclude that if $\sigma_0$ is large and $\sigma$ is small, strong correlation will exhibit between $\alpha_1$ and $\alpha_2$.
Furthermore, consider the case where $\sigma$ is fixed. Hence, if we have a strong prior, \ie, $\sigma_0 \to 0$, then $\alpha_1$ and $\alpha_2$ will show no correlation (and indeed are  independent). If, however, we have a weak prior, \ie, $\sigma_0\to \infty$, then strong correlation exists between $\alpha_1$ and $\alpha_2$.
In MPME, we further exploit the fact that the mean values are close to each other, \ie, $\sigma$ is small. It is evident from \eqnref{co2} that this also advocates strong correlation across different populations.
Similar arguments can be made for the case of multiple populations.

In general, for a graphical model shown in \figref{introgm1} with non-trivial conditional probability distributions, $\alpha_i$'s are correlated \cite{kollar2009probabilistic}.

\section{Learning the uniform prior and MAP using the uniform prior}
\seclabel{unidetails}

\subsection{Learning Hyperparameters using MLE}
According to the graphical model for the UNI prior in \figref{uniform_prior},  the following conditional independence relationships are satisfied
\begin{equation}
\begin{aligned}
(\mu_i &\ci \sigma_i^2 |a,b,c,d), \\
( \mu_1 &\ci \cdots \ci \mu_P | a, b),\\
( \sigma_1^2 &\ci \cdots \ci  \sigma_P^2 | c, d),\\
(\mc{X}_1 &\ci \cdots \ci \mc{X}_P | \vec{\mu}, \vec{\sigma^2}). \\
\end{aligned}
\label{eqn:unici}
\end{equation}
Applying \eqnref{unici}, \eqnref{likelihood0} can be simplified
\begin{equation}
\begin{aligned}
& p(\mc{X}_1, \cdots, \mc{X}_P | \vec{\theta}) \\
=& \int_{\vec{\mu}, \vec{\sigma^2}} p(\mc{X}_1, \cdots, \mc{X}_P | \vec{\mu}, \vec{\sigma^2}) p(\vec{\mu}, \vec{\sigma^2}|\vec{\theta}) d\vec{\mu} d\vec{\sigma^2}\\ 
=& \int_{\vec{\mu}, \vec{\sigma^2}}  \left( \prod_{i=1}^P p(\mc{X}_i |\mu_i, \sigma_i^2) \right) \left( \prod_{i=1}^P p(\mu_i, \sigma_i^2 | \vec{\theta}) \right) d\vec{\mu} d\vec{\sigma^2}\\ 
=&   \prod_{i=1}^P  \int_{\mu_i, \sigma_i^2} p(\mc{X}_i |\mu_i, \sigma_i^2) p(\mu_i, \sigma_i^2 | \vec{\theta}) d\mu_i d\sigma_i^2\\ 
=&   \prod_{i=1}^P  \int_{\mu_i, \sigma_i^2} p(\mc{X}_i |\mu_i, \sigma_i^2) p(\mu_i|a, b) p (\sigma_i^2|c, d) d\mu_i d\sigma_i^2.\\ 
\end{aligned}
\label{eqn:likelihood1}
\end{equation}

For the UNI prior, we have $p(\mu_i, \sigma_i^2 | a, b, c, d) = p(\mu_i|a,b) p(\sigma_i^2 | c,d)$, \ie,
\begin{equation}
p(\mu_i, \sigma_i^2 | \vec{\theta}) = 
 \left\{ 
  \begin{array}{l l}
	\frac{1}{b-a} \frac{1}{d-c}, & \text{if $a\le\mu_i\le b, c\le \sigma_i^2 \le d$,}\\
	0,  & \text{otherwise.}\\
  \end{array} \right.
\label{eqn:pmu}
\end{equation}

Inserting \eqnref{pmu} and \eqnref{px1} into \eqnref{likelihood1}, we need to compute for each $i$, 
\begin{footnotesize}
\begin{equation}
\begin{aligned}
&\int_0^\infty \int_{-\infty}^\infty p(\mc{X}_i | \mu_i, \sigma_i^2) p(\mu_i, \sigma_i | a, b, c, d) d\mu_i d\sigma_i^2\\
\propto &\frac{1}{b-a} \frac{1}{d-c}\times\\
\bigg[ &Q_{N_i-3} \left( \frac{\left(b-\bar{x}_i\right)\sqrt{N_i(N_i-3)}}{\sqrt{(N_i-1)S_i}}, 0; 0, \sqrt{\frac{(N_i-1)S_i}{c}} \right) \\
-&Q_{N_i-3}\left( \frac{\left(a-\bar{x}_i\right)\sqrt{N_i(N_i-3)}}{\sqrt{(N_i-1)S_i}}, 0; 0, \sqrt{\frac{(N_i-1)S_i}{c}} \right) \\
-&Q_{N_i-3}\left( \frac{\left(b-\bar{x}_i\right)\sqrt{N_i(N_i-3)}}{\sqrt{(N_i-1)S_i}}, 0; 0, \sqrt{\frac{(N_i-1)S_i}{d}} \right) \\
-&Q_{N_i-3}\left( \left.\frac{\left(a-\bar{x}_i\right)\sqrt{N_i(N_i-3)}}{\sqrt{(N_i-1)S_i}}, 0; 0, \sqrt{\frac{(N_i-1)S_i}{d}} \right) \right].\\
\end{aligned}
\label{eqn:pjoint2}
\end{equation}
\end{footnotesize}

If $\phi(\cdot)$ and $\Phi(\cdot)$ denotes the PDF and CDF of standard normal distribution respectively, then $Q_f(t,\delta;0,R)$ is defined as:
\begin{equation}
Q_f(t,\delta;0,R) =  \int_0^R \frac{\sqrt{2\pi}y^{f-1}\phi(y)}{\Gamma(\frac{f}{2})2^{\frac{f-2}{2}}} \Phi\left(\frac{ty}{\sqrt{f}}-\delta\right) dy,
\end{equation}
which can be solved by repeated integration by parts to yield closed form solutions \cite{owen1980table}.

\ignore{
where the integral with respect to $\mu_i$ is
\begin{equation}
\begin{aligned}
& \int_a^b p(\mu_i | a, b) d\mu_i p(\mc{X}_i | \mu_i, \sigma_i)  \\
=& \frac{1}{Z} \frac{1}{(b-a)\sqrt{N_i}} \left\{ \Phi( \frac{b - \bar{x}_i}{\sigma_i / \sqrt{N_i}}) - \Phi( \frac{a - \bar{x}_i}{\sigma_i / \sqrt{N_i}})\right\},
\end{aligned}
\end{equation}
where $Z$ is a normalizing constant, and $\Phi(\cdot)$ is the CDF of the standard normal distribution.

(Unfortunately, the integral in terms of $\sigma_i$ is more involved, and we compute it by numerical methods.)
\eqnref{pjoint2} can then be inserted into \eqnref{likelihood1} to compute the likelihood function.
}
\subsection{MAP Estimation}
\label{sec:unimap}
For uniform priors of $\mu_i$ and $\sigma_i^2$, the right-hand side of \eqnref{post1} is
\begin{equation}
\frac{1}{b-a} \frac{1}{d-c} p(\mc{X}_i | \mu_i,\sigma_i^2), \quad \text{if $\mu_i \in [a,b]$ and $\sigma_i^2 \in [c,d]$}.
\end{equation}
Therefore, MAP is equivalent to maximum likelihood estimation on the support $\mu_i \in [a,b]$ and $\sigma_i^2 \in [c,d]$. The solution is simply
\begin{equation}
\mu_{i,MAP} = \left\{ 
  \begin{array}{l l}
	a & \quad \text{if $\mu_{i,MLE}<a$}\\
	\mu_{i,MLE} & \quad \text{if $a\le \mu_{i,MLE}\le b$}\\
	b & \quad \text{if $\mu_{i,MLE}>b$}\\
  \end{array} \right. ,
\label{eqn:mapmeanuni}
\end{equation}
\begin{equation}
\sigma_{i,MAP}^2 = \left\{ 
  \begin{array}{l l}
	c & \quad \text{if $\sigma_{i,MLE}<c$}\\
	\sigma_{i,MLE}^2 & \quad \text{if $c\le \sigma_{i,MLE}\le d$}\\
	d & \quad \text{if $\sigma_{i,MLE}>d$}\\
  \end{array} \right. ,
\label{eqn:mapsigma}
\end{equation}
where $\mu_{i,MLE}$ and $\sigma_{i,MLE}^2$\footnote{$\sigma_{i,MLE}^2$ is a biased estimator. To eliminate the bias, we may replace $\sigma_{i,MLE}^2$ in \eqnref{mapsigma} by its unbiased estimator.} are equal to the sample mean and sample variance estimators, respectively\cite{bookprob1}.

\section{Learning the NIX prior and MAP using the NIX prior}
\seclabel{nixdetails}

\subsection{Learning Hyperparameters using MLE}
According to the graphical model for the NIX prior in \figref{normal_inv_chis_prior},  the following conditional independence relationships are satisfied
\begin{equation}
\begin{aligned}
( \sigma_1^2 \ci \cdots \ci  \sigma_P^2 &| \nu_0, \sigma_0^2),\\
( \mu_1 \ci \cdots \ci \mu_P &| \kappa_0, \mu_0, \nu_0, \sigma_0^2),\\
( \mc{X}_1 \ci \cdots \ci \mc{X}_P &| \vec{\mu}, \vec{\sigma^2}). \\
\end{aligned}
\label{eqn:nixci}
\end{equation}

Applying \eqnref{nixci}, \eqnref{likelihood0} can be simplified
\begin{equation}
\begin{aligned}
& p(\mc{X}_1, \cdots, \mc{X}_P | \vec{\theta}) \\
=& \int_{\vec{\mu}, \vec{\sigma^2}} p(\mc{X}_1, \cdots, \mc{X}_P | \vec{\mu}, \vec{\sigma^2}) p(\vec{\mu}, \vec{\sigma^2}|\vec{\theta}) d\vec{\mu} d\vec{\sigma^2}\\ 
=& \int_{\vec{\mu}, \vec{\sigma^2}}  \left( \prod_{i=1}^P p(\mc{X}_i |\mu_i, \sigma_i^2) \right) \left( \prod_{i=1}^P p(\mu_i, \sigma_i^2 | \vec{\theta}) \right) d\vec{\mu} d\vec{\sigma^2}\\ 
=&   \prod_{i=1}^P  \int_{\mu_i, \sigma_i^2} p(\mc{X}_i |\mu_i, \sigma_i^2) p(\mu_i, \sigma_i^2 | \vec{\theta}) d\mu_i d\sigma_i^2.\\  
\end{aligned}
\label{eqn:likelihood2}
\end{equation}

For the NIX prior, we have $p(\mu_i, \sigma_i^2 |\kappa_0, \mu_0, \nu_0, \sigma_0^2) = p(\sigma_i^2 | \nu_0, \sigma_0^2) p(\mu_i|\sigma_i^2, \mu_0, \kappa_0,)$, \ie,
\begin{equation}
\begin{aligned}
&p(\mu_i, \sigma_i^2 | \vec{\theta})\\
=& \mc{N}(\mu_i | \mu_0, \sigma_i^2 / \kappa_0)\chi^{-2}(\sigma_i^2 | \nu_0, \sigma_0^2) \\
=&\frac{\sigma_i^{-\nu_0-3}}{Z(\kappa_0, \mu_0, \nu_0, \sigma_0^2)}\exp \left\{ -\frac{\nu_0\sigma_0^2+\kappa_0(\mu-\mu_0)^2}{2\sigma_i^2} \right\},\\
\end{aligned}
\label{eqn:pnix}
\end{equation}
where $Z(\kappa_0, \mu_0, \nu_0, \sigma_0^2)$ is a normalizing constant depending on the hyperparameters, explicitly
\begin{equation}
Z(\kappa_0, \mu_0, \nu_0, \sigma_0^2) = \sqrt{\frac{2\pi}{\kappa_0}}\Gamma\left(\frac{\nu_0}{2}\right)\left(\frac{2}{\nu_0\sigma_0^2}\right)^{\nu_0/2}.
\label{eqn:nixz}
\end{equation}

\begin{figure*}
\footnotesize
\begin{equation}
\begin{aligned}
&\int_0^\infty\int_{-\infty}^\infty p(\mc{X}_i | \mu_i, \sigma_i^2) p(\mu_i, \sigma_i | \kappa_0, \mu_0, \nu_0, \sigma_0^2) d\mu_i d\sigma_i^2\\
&=\int_0^\infty\int_{-\infty}^\infty \frac{\sigma_i^{-N_i}}{(2\pi)^{N_i/2}}\exp \left\{-\frac{N_i(\bar{x}_i-\mu_i)^2+(N_i-1)S_i}{2\sigma_i^2} \right\}\frac{\sigma_i^{-\nu_0-3}}{Z(\kappa_0, \mu_0, \nu_0, \sigma_0^2)}\exp \left\{-\frac{\nu_0\sigma_0^2+\kappa_0(\mu-\mu_0)^2}{2\sigma_i^2} \right\} d\mu_i d\sigma_i^2\\
&=\frac{1}{(2\pi)^{N_i/2}}\frac{1}{Z(\kappa_0, \mu_0, \nu_0, \sigma_0^2)}\int_0^\infty\int_{-\infty}^\infty \sigma_i^{-\nu_{N_i,i}-3} \exp \left\{ -\frac{\nu_{N_i,i}\sigma_{N_i,i}^2+\kappa_{N_i,i}(\mu-\mu_{N_i,i})^2}{2\sigma_i^2} \right\} d\mu_i d\sigma_i^2\\
&=\frac{1}{(2\pi)^{N_i/2}}\frac{Z(\kappa_{N_i,i}, \mu_{N_i,i}, \nu_{N_i,i}, \sigma_{N_i,i}^2)}{Z(\kappa_0, \mu_0, \nu_0, \sigma_0^2)}\int_0^\infty\int_{-\infty}^\infty \mc{N}(\mu_i | \mu_{N_i,i}, \sigma_i^2 / \kappa_{N_i,i})\chi^{-2}(\sigma_i^2 | \nu_{N_i,i}, \sigma_{N_i,i}^2) d\mu_i d\sigma_i^2\\
&=\frac{1}{(2\pi)^{N_i/2}}\frac{Z(\kappa_{N_i,i}, \mu_{N_i,i}, \nu_{N_i,i}, \sigma_{N_i,i}^2)}{Z(\kappa_0, \mu_0, \nu_0, \sigma_0^2)}.\\
\end{aligned}
\label{eqn:margnix}
\end{equation}
\hrulefill
\vspace*{4pt}
\end{figure*}

Inserting \eqnref{pnix} and \eqnref{px1} into \eqnref{likelihood2}, we need to compute for each $i$ as given in \eqnref{margnix} where $\kappa_{N_i,i}, \mu_{N_i,i}, \nu_{N_i,i}$ and $\sigma_{N_i,i}^2$ are defined as
\begin{equation}
\begin{aligned}
\kappa_{N_i,i} &= \kappa_0+N_i,\\
\mu_{N_i,i} &= \frac{\kappa_0\mu_0+N_i\bar{x}_i}{\kappa_{N_i,i}},\\
\nu_{N_i,i} &= \nu_0+N_i,\\
\sigma_{N_i,i}^2 &= \frac{\nu_0}{\nu_{N_i,i}}\sigma_0^2+\frac{N_i-1}{\nu_{N_i,i}}S_i+\frac{\kappa_0 N_i(\mu_0-\bar{x}_i)^2}{\nu_{N_i,i}\kappa_{N_i,i}}.\\
\end{aligned}
\end{equation}

Substituting \eqnref{nixz} into \eqnref{margnix} and then back into \eqnref{likelihood2}, we obtain the likelihood in closed form as
\begin{equation}
\begin{aligned}
&p(\mc{X}_1, \cdots, \mc{X}_P | \mu_0, \kappa_0, \nu_0, \sigma_0^2)\\
&=\prod_{i=1}^P \frac{\Gamma(\nu_{N_i,i}/2)}{\Gamma(\nu_0/2)}\sqrt{\frac{\kappa_0}{\kappa_{N_i,i}}}\frac{(\nu_0\sigma_0^2)^{\nu_0/2}}{(\nu_{N_i,i}\sigma_{N_i,i}^2)^{\nu_{N_i,i}/2}}\frac{1}{\pi^{N_i/2}}.
\end{aligned}
\end{equation}

\subsection{MAP Estimation}
\label{sec:nixmap}
For NIX priors of $\mu_i$ and $\sigma_i$, the posterior in \eqnref{map} is
\begin{equation}
\mc{N}(\mu_i | \mu_{N_i,i}, \sigma_i^2 / \kappa_{N_i,i})\chi^{-2}(\sigma_i^2 | \nu_{N_i,i}, \sigma_{N_i,i}^2).
\end{equation}
Therefore, MAP estimates of $\mu_i$ and $\sigma_i$ are the modes of the posterior, which can be seen to be simply \cite{murphy2012machine}
\begin{equation}
\begin{aligned}
\mu_{i,MAP} &= \mu_{N_i,i}= \frac{\kappa_0\mu_0 + \sum_{j=1}^{N_i}x_{ij}}{\kappa_0+N_i}\\
\sigma_{i,MAP}^2 &= \frac{\nu_{N_i,i}\sigma_{N_i,i}^2}{\nu_{N_i,i}+3}.
\end{aligned}
\label{eqn:mapsigma2}
\end{equation}

The simplified expression for $\mu_{i,MAP}$ can be interpreted as adding $\kappa_0$ number of fake data samples with value (and so mean also) $\mu_0$ to the measured data $\mc{X}_i$ for population $i$.  Similarly by expanding expression for $\sigma_{i,MAP}^2$ one can obtain an analogous elucidation for the variance update. Basically this additional fake data samples incorporates the information present in the prior learnt using data from all the populations.

Furthermore, note that $\sigma_{i,MAP}^2$ in \eqnref{mapsigma2} is a biased estimator. An unbiased estimation can be obtained by
\begin{equation}
\begin{aligned}
\sigma_{i,MAP-UB}^2 &= \frac{\nu_{N_i,i}\sigma_{N_i,i}^2}{\nu_{N_i,i}-1}.
\end{aligned}
\label{eqn:mapsigma3}
\end{equation}

\section*{Acknowledgment}

The authors would like to thank Eli Chiprout for various discussions and continuous support.
This work is in part supported by Intel Corporation.

\bibliographystyle{IEEEtran}
\bibliography{mpme-v20}

%




\end{document}